\documentclass{jfm}
\usepackage{graphicx,float,rotating,multirow}
\usepackage{amsfonts,amssymb,latexsym,color,amsmath,pifont,epsfig,mathdots,mathtools,natbib,cancel}
\usepackage{setspace,subfigure,url,epstopdf}

\providecommand{\bnabla}{\boldsymbol{\nabla}}

\newcommand{\Ret}{Re_\tau}  

\newcommand{\eg}{e.g.\,}
\newcommand{\ie}{i.e.\,}

\newcommand{\cL}{\mathbf{\cal L}}
\newcommand{\cH}{\mathbf{\cal \tilde{H}}}

\newcommand{\fb}{\mathbf{f}}
\newcommand{\ub}{\mathbf{u}}
\newcommand{\kb}{\mathbf{k}}

\newcommand{\uk}{\mathbf{u_k}}
\newcommand{\uo}{\mathbf{u_0}}

\newcommand{\fk}{\mathbf{f_k}}
\newcommand{\pk}{p_\mathbf{k}}
\newcommand{\sk}{\sigma_{\mathbf{k}}}
\newcommand{\sko}{\sigma_{\mathbf{k}0}}
\newcommand{\skc}{\sigma_{\mathbf{k}c}}

\newcommand{\kx}{\kappa_x}
\newcommand{\kz}{\kappa_z}
\newcommand{\om}{\omega}

\newcommand{\lxp}{\lambda^+_x}
\newcommand{\lzp}{\lambda^+_z}
\newcommand{\ypc}{y^+_c}

\newcommand{\p}{\partial}

\title{A framework for studying the effect of compliant surfaces on wall turbulence}
\shorttitle{Wall turbulence over compliant surfaces}

\author{M. Luhar$^{1}$\thanks{Current affiliation: Department of Aerospace and Mechanical Engineering, University of Southern California, CA 90089, USA.  Email address for correspondence: luhar@usc.edu}, 
	A. S. Sharma$^{2}$ \and \ B. J. McKeon$^{1}$}

\affiliation{
	$^{1}$Graduate Aerospace Laboratories, California Institute of Technology, CA 91125, USA \\[0.1cm]
	$^{2}$Engineering and the Environment, University of Southampton, SO17 1BJ, UK \\[0.1cm]}

\shortauthor{M. Luhar, A. S. Sharma, and B. J. McKeon}

\date{?; revised ?; accepted ?. - To be entered by editorial office}

\begin{document}
\maketitle
\begin{abstract}
This paper extends the resolvent formulation proposed by \citet{McKeon2010} to consider turbulence--compliant wall interactions.  Under this formulation, the turbulent velocity field is expressed as a linear superposition of propagating modes, identified via a gain-based decomposition of the Navier--Stokes equations.  Compliant surfaces, modeled as a complex wall admittance linking pressure and velocity, affect the gain and structure of these modes.  With minimal computation, this framework accurately predicts the emergence of the quasi-2D propagating waves observed in recent direct numerical simulations.  Further, the analysis also enables the rational design of compliant surfaces, with properties optimized to suppress flow structures energetic in wall turbulence.  It is shown that walls with unphysical negative damping are required to interact favorably with modes resembling the energetic near-wall cycle, which could explain why previous studies have met with limited success.  Positive-damping walls are effective for modes resembling the so-called very large-scale motions (VLSMs), indicating that compliant surfaces may be better suited for application at higher Reynolds number.  Unfortunately, walls that suppress structures energetic in natural turbulence are also predicted to have detrimental effects elsewhere in spectral space.  Consistent with previous experiments and simulations, slow-moving spanwise-constant structures are particularly susceptible to further amplification.  Mitigating these adverse effects will be central to the development of compliant coatings that have a net positive influence on the flow.
\end{abstract}

\section{Introduction}\label{sec:intro}
It has long been recognized that compliant walls have the potential to serve as passive controllers for turbulent flows \citep{GadelHak2000}.  The fact that such walls would require no energy input for control and involve no sensors, actuators or complex control algorithms makes them especially attractive from a practical standpoint.  However, despite numerous efforts since the early experiments of \citet{Kramer1961}, designing a compliant surface that can interact favorably with wall-bounded turbulent flows and reduce skin friction drag remains an open challenge.

Classical instability analyses have provided significant insight into the possibility of transition delay using compliant walls \citep{Benjamin1960,Landahl1962,Carpenter1985,Carpenter1986,Lucey1995,Davies1997}, which has led to an elegant energy-based classification of the various fluid and structural instabilities possible \citep{Benjamin1963}.  Some studies suggest that the use of multiple compliant panels of different material properties may even delay transition indefinitely \citep{Carpenter2000}.  Yet, to our knowledge, there is no universally accepted theoretical framework to study the interaction between \textit{turbulent} flows and compliant surfaces.  Without this guiding framework, time and resource intensive experiments \citep[\eg][]{Bushnell1977,Lee1993,Choi1997} and direct numerical simulations \citep[DNS, \eg][]{Endo2002,Xu2003,Fukagata2008,Kim2014} have been restricted to relatively small parameter ranges and met with limited success.

\subsection{Previous studies}\label{sec:intro-Previous}
For a critical review of the early experimental efforts studying the interaction between compliant walls and turbulent boundary layers, the reader is referred to \citet{Bushnell1977}.  A comprehensive review of the theoretical and experimental efforts after 1977 can be found in \citet{Riley1988} and \citet{Carpenter1990}.  Since then, the hydrogen bubble flow-visualization experiments pursued by \citet{Lee1993} have shown that a single-layer viscoelastic coating can lead to an intermittent relaminarization-like phenomenon.  These experiments did not yield skin friction estimates but \cite{Lee1993} did report a reduction in the Reynolds stress and streamwise energy intensity over the compliant surface.  Subsequent water-tunnel experiments by \citet{Choi1997} have confirmed that a viscoelastic coating made of silicon rubber can reduce the skin friction by $3-7\%$ at low free-stream velocities.  However, these promising results should be treated with a degree of caution since the experimental error associated with the skin friction measurements was estimated to be as much as $\pm 4\%$ and another coating with a larger loss coefficient (\ie greater viscous damping) resulted in a skin friction increase.

While the experiments provide some cause for optimism, results from numerical simulations have been inconclusive.  The compliant wall tested by \citet{Xu2003} in their DNS study led to very little modification of the near-wall coherent structures and no change in the long-time skin friction drag.  The optimization performed by \citet{Fukagata2008} showed that anisotropic compliant surfaces could yield skin friction reductions of up to $8\%$ for very low Reynolds number turbulent channel flow (bulk Reynolds number $Re_b = 3300$).  However, performance deteriorated as the size of the flow domain increased.  A recent DNS study by \citet{Kim2014} showed very little change in skin-friction over stiff walls (characterized by a high spring constant, damping and tension).  Softer walls led to the generation of large-amplitude two-dimensional traveling waves that resulted in a significant increase in drag.

There have also been some previous attempts at developing simplified models for the study of turbulence--compliant wall interactions.  \citet{Duncan1986} developed a two-dimensional, quasi-interactive scheme wherein an unsteady pressure pulse of an assumed shape was imposed on a compliant surface of known material properties.  The interaction between the deflecting compliant wall and a mean potential flow was then computed explicitly.  \citet{Kireiko1990} used a linearized monoharmonic approximation to consider the interaction between near-wall turbulence and compliant surfaces.  

More recently, \citet{Rempfer2001} developed a low-order dynamical model for the fluid-structure system that employed a Galerkin projection onto a combination of (i) basis functions obtained via proper orthogonal decomposition (POD) of a rigid-wall DNS data set and (ii) Stokes eigenfunctions to account for the compliant wall boundary condition.  The wall itself was assumed to act as a simple second-order mass-spring-damper system (although it is clear that the model can be extended to account for more complex surfaces).  This last study by \citet{Rempfer2001} provides a reasonably complete framework for the study of turbulence--compliant wall interactions.  However, it does have some limitations.  For example, the requirement of a preexisting DNS database for the POD restricts the model to low Reynolds numbers.  \citet{Rempfer2001} also noted that the \textit{ad hoc} use of Stokes eigenfunctions to satisfy the boundary conditions may not adequately describe the near-wall physics.

\subsection{Contribution}\label{sec:intro-Contribution}
The lack of any definitive results highlights the need for a computationally inexpensive framework that allows an exploration of the parameter space and enables the rational design of performance enhancing compliant walls.  In an effort to address this need, this paper extends the resolvent analysis of \citet{McKeon2010}.  Under the resolvent formulation, the turbulent velocity and pressure fields are represented by a series of highly-amplified flow structures identified directly from the governing Navier--Stokes equations (NSE) through a gain-based decomposition.  Specifically, this analysis interprets the Fourier-transformed NSE as a forcing-response system with feedback; the nonlinear convective terms are treated as the forcing that is processed by the linear terms to generate a velocity and pressure response.  A singular value decomposition (SVD) of the forcing-response transfer function --- the resolvent operator --- is used to identify the most amplified velocity and pressure fields at each wavenumber-frequency combination.  These high-gain traveling-wave structures, termed resolvent modes, are assumed to dominate the flow field.

Subsequent studies have shown that the resolvent modes yield predictions consistent with the structure, statistics, and scaling of wall-bounded turbulent flows.  For example, certain resolvent modes reproduce features of the dynamically important near-wall cycle (NW) while others resemble the so-called very large-scale motions (VLSMs) that appear in the logarithmic region at higher Reynolds numbers and have an organizing influence on the near-wall flow  \citep{McKeon2013}.  \citet{Sharma2013} demonstrated that some modes naturally give rise to hairpin-like vortices and that a small number of modes can be combined to yield realistic structure resembling modulating hairpin packets.  \citet{Moarref2013} identified different scaling regimes for the resolvent modes as a function of their wave speed and used this knowledge to develop a model for the streamwise energy intensity profile.

For the purposes of the present study, two further developments are key.  First, \citet{Luhar2014b} recently showed that resolvent modes reconcile many of the key relationships between the velocity field, wall-pressure fluctuations, and coherent structure observed in previous experiment and DNS.  Second, using opposition control \citep{Choi1994} as an example, \citet{Luhar2014a} demonstrated that these modes also serve as effective building blocks for reduced-order models that can be used to evaluate and design flow control schemes.  Thus, the resolvent modes can reasonably represent turbulent flow structures, capture their wall-pressure footprint, and account for the effects of control.  Motivated by these developments, the present study employs them to consider the effect of compliant surfaces.  The compliant wall is introduced via a change in the dynamic boundary conditions for the NSE (\ie the pressure-velocity relationship at the wall) prior to performing the SVD.  The resulting change in the structure and amplification of the resolvent modes relative to the rigid-wall case is used to infer the effectiveness of the compliant surface.

Note that the framework developed in this paper is conceptually similar to the work of \citet{Rempfer2001} in that it considers the effect of compliant walls on a limited set of basis functions (resolvent modes).  However, since the rigid-wall (null-case) and compliant-wall resolvent modes are identified directly from the governing equations, there are no Reynolds number limitations for the present study and the boundary conditions are satisfied automatically.  Another advantage of the present approach is that is permits a mode-by-mode evaluation of control.  This is computationally inexpensive, which makes it possible to perform searches for optimal wall properties that are tuned to suppress structures energetic in natural flows.  As will be shown later, such blind searches occasionally identify non-traditional wall properties to be optimal.  This finding in itself is important because it indicates that designing effective compliant walls might call for the use of \textit{metamaterials} or require further advances in materials development.

\section{Approach}\label{sec:model}
This section provides a brief review of the resolvent formulation proposed by \citet{McKeon2010} before presenting the extension to account for compliant walls.  To maintain consistency with DNS studies considering compliant surfaces \citep{Xu2003,Fukagata2008,Kim2014}, this paper only considers fully-developed channel flow.  However, the approach presented herein can be extended to turbulent pipe and boundary layer flows as well.

\subsection{Resolvent analysis}\label{sec:model-Resolvent}
As discussed in \citet{McKeon2010} and \citet{McKeon2013}, the resolvent formulation stems from the insight that the governing NSE can be considered a forcing-response system with feedback.  The conservative nonlinear terms, which serve to transfer energy across spectral space, are interpreted as the feedback forcing that acts on the linear terms of the NSE to generate a velocity and pressure response.  Importantly, it turns out that the linear terms behave like highly directional filters and so the flow response is (almost) insensitive to the nature of the nonlinear forcing.  \citet{McKeon2013} argue that this directionality is partially responsible for the robust nature of wall turbulence, \ie the presence of repeating, recognizable flow patterns across geometries, scales and operating conditions (NW cycle, hairpins, VLSMs).  Given this directionality, a gain-based decomposition of the forcing-response transfer function is used to arrive at a low-order representation for the flow field.

For fully developed turbulent channel flow, spatial invariance and statistical stationarity ensure that Fourier modes are the most appropriate bases for decomposition in the streamwise direction $x$, spanwise direction $z$, and time $t$:

\begin{align}\label{eqFourier}
\left[\begin{array}{c} \ub(x,y,z,t) \\ p(x,y,z,t) \end{array}\right] 
= \iiint_{-\infty}^{\infty} \left[\begin{array}{c} \uk(y) \\ \pk(y) \end{array}\right] e^{i(\kx x + \kz z - \om t)} 
d\kx d\kz d\om.
\end{align}

{\noindent}Here, $\ub = [u , v , w]^T$ represents the streamwise ($u$), wall-normal ($v$) and spanwise ($w$) velocity fields and $p$ denotes the pressure field.  Each wavenumber-frequency combination $\kb = (\kx,\kz,\om)$ represents a flow structure, or mode, with streamwise and spanwise wavelengths $\lambda_x = 2\pi/\kx$ and $\lambda_z = 2\pi/\kz$ propagating downstream at speed $c = \om/\kx$.  (\textit{n.b.} unless otherwise indicated, length scales are normalized with respect to the channel half-height $h$ and velocity scales are normalized with respect to the friction velocity $u_\tau$.  Following standard notation, a superscript $+$ is used to denote normalization with respect to $u_\tau$ and the kinematic viscosity $\nu$).  The Fourier coefficients $\uk$ and $\pk$ represent the wall-normal ($y$) variation in the magnitude and phase of the velocity and pressure fields for each mode.  The special case $\uo = [U(y) , 0 , 0]^T$ represents the mean velocity profile.

For the inhomogeneous wall-normal direction, the analysis seeks a gain-based decomposition.  Specifically, the Fourier-transformed NSE and continuity constraint are expressed as the following forcing-response system:

\begin{align}\label{eqForcingResponse}
\left[\begin{array}{c} \uk \\ \pk \end{array}\right] 
&= 
\left( -i\omega \left[\begin{array}{cc} \mathbf{I}  &  \\  &0\\ \end{array}\right] -
\left[\begin{array}{cc} \cL_\kb & -\nabla_\kb\\ \nabla_\kb^T & 0\\ \end{array}\right] \right)^{-1}
\left[\begin{array}{cc} \mathbf{I}\\0\\ \end{array}\right] \fk \nonumber \\ 
& \nonumber \\
&= \cH_\kb \fk,
\end{align}

{\noindent}where $\nabla_\kb = [i\kx , \p/\p y , i\kz]^T$ and $\nabla_\kb^T$ represent the gradient and divergence operators, respectively, and $\fk = (-\ub \cdot \bnabla \ub)_\kb$ represents the nonlinear forcing.  The forcing is mapped to a velocity and pressure response by the resolvent $\cH_\kb$, which depends on the linear operator:

\begin{align}\label{eqLinearOperator}
\cL_\kb
&= 
\left[\begin{array}{ccc} 
-i\kx U + \Ret^{-1} \nabla_\kb^2  & -\p U/\p y & 0 \\
0 & -i\kx U + \Ret^{-1} \nabla_\kb^2 & 0 \\
0 & 0 & -i\kx U + \Ret^{-1} \nabla_\kb^2 \\
\end{array}\right].
\end{align}

{\noindent}Here, $\Ret = u_\tau h / \nu$ is the friction Reynolds number and $\nabla_\kb^2 = [-\kx^2 + \p^2/\p y^2 - \kz^2]$ is the Fourier-transformed Laplacian.  The first line in the operator on the right-hand side of (\ref{eqForcingResponse}) represents the momentum equations and the second line represents the continuity constraint.

Following \citet{McKeon2010}, a singular value decomposition (SVD) of the resolvent operator $\cH_\kb$, discretized using a Chebyshev collocation method (\S \ref{sec:model-Validation}), yields a set of orthonormal forcing ($\fb_{\kb,m}$) and response ($[\ub_{\kb,m},p_{\kb,m}]^T$) modes, ordered based on the input-output gain ($\sigma_{\kb,m}$) under an $L^2$ energy norm.  To enforce this energy norm, the primitive-variable resolvent in (\ref{eqForcingResponse}) is scaled as:

\begin{gather}
\left[\begin{array}{cc} W_\ub^{} & 0 \\ \end{array}\right] \left[\begin{array}{c} \uk \\ p_\kb \end{array}\right]
= \left( \left[\begin{array}{cc} W_\ub^{} & 0 \\ \end{array}\right] \cH_\kb W_\fb^{-1} \right) W_\fb^{} \fk
\end{gather}

{\noindent}or

\begin{gather}
W_\ub^{} \uk = \cH_\kb^S W_\fb^{} \fk.
\label{eqScaledH}
\end{gather}

{\noindent}Note that the discretized resolvent operator $\cH_\kb$ is a block matrix of dimension $[4 \times 3]$ while the scaled resolvent operator $\cH_\kb^S$ is of dimension $[3 \times 3]$.  The $[3 \times 3]$ block diagonal matrices $W_\ub$ and $W_\fb$ incorporate numerical quadrature weights which ensure that the SVD of the scaled resolvent operator:

\begin{gather}
\cH_\kb^S = \sum\limits_{m} \psi_{\kb,m} \sigma_{\kb,m} {\phi^*_{\kb,m}},
\label{eqScaledSVD}
\end{gather}

{\noindent}where

\begin{gather}
\sigma_{\kb,1} > \sigma_{\kb,2}...  > \sigma_{\kb,m}... >0, \:\:
{\phi^*_{\kb,l}}{\phi^{}_{\kb,m}} = \delta_{lm}, \:\: 
{\psi^*_{\kb,l}}{\psi^{}_{\kb,m}} = \delta_{lm},
\label{eqSVD}
\end{gather}

{\noindent}yields forcing modes ${\fb_{\kb,m}} = W_\fb^{-1} {\phi_{\kb,m}}$ and velocity response modes ${\ub_{\kb,m}} = W_\ub^{-1} {\psi_{\kb,m}}$ with unit energy over the channel cross-section, \ie the orthonormality constraints (\ref{eqSVD}) translate into

\begin{equation}\label{eqSVDNorm}
\int \fb^*_{\kb,l} \fb^{}_{\kb,m} \,dy = \delta_{lm} \:\: , \:\: \int \ub^*_{\kb,l} \ub^{}_{\kb,m} \,dy = \delta_{lm},
\end{equation}

{\noindent}where a superscript $*$ denotes a complex conjugate.

From (\ref{eqForcingResponse}-\ref{eqSVD}), it can be seen that forcing in the direction of the $m^{th}$ singular forcing mode with unit amplitude results in a response in the direction of the $m^{th}$ singular response mode amplified by the singular value $\sigma_{\kb,m}$, \ie forcing $\fk=\fb_{\kb,1}$ creates a response $[\uk,\pk]^T = \sigma_{\kb,1} [\ub_{\kb,1},p_{\kb,1}]^T$.  \citet{Luhar2014b} provide a more detailed description of how the primitive variable resolvent shown in (\ref{eqForcingResponse}) is scaled to enforce the energy norm and how the pressure field is extracted from the unscaled operator.

Importantly, for $\kb$ combinations energetic in natural turbulence, the resolvent operator tends to be low rank.  Only a limited number of inputs are highly amplified.  Frequently, $\sigma_{\kb,1} \gg \sigma_{\kb,2}$ and so the velocity and pressure fields can be reasonably approximated by the first response modes, $\ub_{\kb,1}$ and $p_{\kb,1}$, respectively.  Recent studies \citep{McKeon2010,McKeon2013,Moarref2013,Luhar2014b} have shown that models based on this rank-1 approximation yield statistical and structural predictions consistent with previous observations for wall turbulence.  The reader is referred to these studies for an expanded discussion of the amplification characteristics of the resolvent operator and the rank-1 approximation.

Given the success of the rank-1 response modes in reproducing previous observations, the present paper employs them for the study of turbulence--compliant wall interactions.  For the remainder of this paper, it is assumed that the velocity and pressure fields at each $\kb$ correspond to the rank-1 responses: $[\uk(y),\pk(y)]^T = [\ub_{\kb,1},p_{\kb,1}]^T$.  The term resolvent modes refers to these rank-1 velocity and pressure fields, while the terms amplification and gain are used interchangeably to refer to the corresponding singular values, $\sigma_{\kb,1}$.  The additional subscript $1$ is dropped for notational convenience.

Assuming that the efficacy of any control scheme depends on its ability to suppress the generation of Reynolds stress \citep{Fukagata2002}, the forcing-response interpretation of the NSE suggests three possible mechanisms: (i) a reduction in the magnitude of the nonlinear forcing $\fk$ present in the flow or a change in the form of the forcing (\eg toward lower-gain directions) such that the magnitude of the velocity response diminishes, (ii) a modification of the system such that the forcing-response gain diminishes, or (iii) a reduction in the Reynolds stress contribution from high-gain resolvent modes via a change in mode structure.  While mechanism (i) clearly requires knowledge of the nonlinear coupling between resolvent modes, mechanisms (ii) and (iii) can be evaluated on a linear mode-by-mode basis.  This mode-by-mode approach is adopted here.

In other words, the present paper only considers the effect of compliant walls on the shape ($\uk$, $\pk$) and forcing-response amplification ($\sk$) of the resolvent modes.  It provides no information on the nonlinear interaction between resolvent modes, which would feedback via the quadratic nonlinearity to force the linearized system.  A more complete model would require knowledge of the triadic coupling between all resolvent modes, which is outside the scope of this paper.  However, even without this knowledge, the change in the gain and structure of the modes can provide significant insight into the problem, as evidenced by the success of the analogous rank-1 model for opposition control developed by \citet{Luhar2014a}.

In addition to the nonlinearity, the mean velocity profile $U(y)$ in (\ref{eqLinearOperator}) is another point of connection between modes.  Here, the discretized resolvent operator is constructed using a mean profile computed based on a turbulent viscosity model \citep[][see \S\ref{sec:model-Validation}]{Reynolds1967}.  If the compliant wall has a significant influence on the flow, the mean velocity profile would no longer resemble the canonical form, which could lead to changes in the gain and structure of the resolvent modes.  Again, in a complete model, the mean profile must be sustained by the total Reynolds stress generated by the resolvent modes to account for this feedback.

\subsection{Compliant wall effects}\label{sec:model-CompliantWall}
For the rigid-wall null case, the discretized resolvent operator in (\ref{eqForcingResponse}) is constructed using the standard no-slip boundary conditions at the lower ($y=0$) and upper walls ($y=2$):

\begin{equation}\label{eqNoSlip}
\uk(0)=\uk(2)=0.
\end{equation}

{\noindent}The effect of the compliant wall is introduced by changing the boundary conditions on velocity and pressure within the resolvent before computing the SVD (\ref{eqSVD}).  Specifically, for wall displacement $\eta(x,z,t)$ constrained to be in the vertical direction, the kinematic boundary conditions at the lower wall, $\ub(\eta) = \p \eta / \p t$, can be expressed as the following Fourier-transformed, linearized Taylor series expansions:

\begin{align}\label{eqKinematicBCu}
u_\kb(\eta) & \approx u_\kb(0) + \eta_\kb \frac{\p U}{\p y}\Big|_{0}
+& \sum\limits_{\kb = \kb_a - \kb_b} \bcancel{\eta_{\kb_a} \frac{\p u^*_{\kb_b}}{\p y}\Big|_{0}} + ...  &=& 0, \\ \label{eqKinematicBCv}
v_\kb(\eta) & \approx v_\kb(0)
+& \sum\limits_{\kb = \kb_a - \kb_b} \bcancel{\eta_{\kb_a} \frac{\p v^*_{\kb_b}}{\p y}\Big|_{0}} + ... &=& -i\om \eta_\kb, \\ \label{eqKinematicBCw}
w_\kb(\eta) & \approx w_\kb(0) 
+& \sum\limits_{\kb = \kb_a - \kb_b} \bcancel{\eta_{\kb_a} \frac{\p w^*_{\kb_b}}{\p y}\Big|_{0}} + ... &=& 0,
\end{align}

{\noindent}where $\eta_\kb$ represents the Fourier coefficient for the wall displacement at wavenumber-frequency combination $\kb = (\kx,\kx,\om)$.  The neglected quadratic terms are shown for reference.   Physically, the kinematic boundary condition (\ref{eqKinematicBCu}) estimates the streamwise velocity at $y=0$ by trying to satisfy the no-slip condition at the wall $y=\eta_\kb$.  This means that the streamwise fluctuations must counter the effects of the mean flow, leading to the additional term $\eta_\kb \left(\p U /\p y \right)_0$.  As will be shown later, this can result in relatively large magnitudes for $u_\kb(0)$.  Further, (\ref{eqKinematicBCu}) and (\ref{eqKinematicBCv}) can be combined to yield the following expression: $i\om u_\kb(0) \approx v_\kb(0) \left(\p U /\p y \right)_0$.  This means that the streamwise and wall-normal velocities have to be $\pi/2$ out of phase near the wall.

The dynamic boundary condition is expressed as a complex mechanical admittance $Y$ linking wall-normal velocity and pressure:

\begin{equation}\label{eqDynamicBC}
Y = \frac{v_\kb(0)}{p_\kb(0)}.
\end{equation}

{\noindent}Similar boundary conditions apply at the top wall, $y=2$.  Although, for walls with identical material properties, the sign of the admittance changes due to the differing symmetry of the pressure and wall-normal velocity fields with respect to the centerline, \ie $v_\kb(0)/p_\kb(0) = -v_\kb(2)/p_\kb(2)$.

\begin{figure}
	\centering
	\includegraphics[width=10cm]{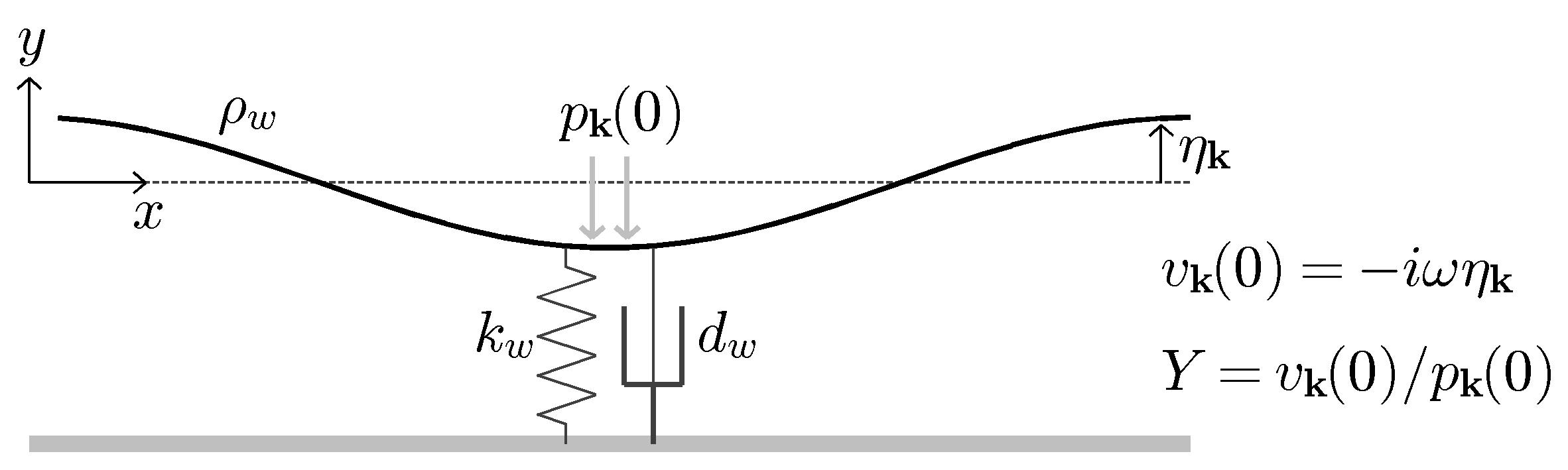}
	\caption{Schematic showing a simple spring- and damper-backed compliant membrane and boundary conditions.}
	\label{fig1:schematic}
\end{figure}

Clearly, the linearization of the kinematic boundary conditions (\ref{eqKinematicBCu}-\ref{eqKinematicBCw}) is a significant simplification since the neglected higher-order terms could be important for large wall displacements.  However, retaining terms of quadratic or higher order in the fluctuations would again require a coupled nonlinear model allowing for interactions between resolvent modes, which is outside the scope of this paper as noted earlier.  Further, the linearized boundary conditions require an estimate of the mean shear at the wall (\ref{eqKinematicBCu}), which is again assumed to correspond to the rigid wall case \citep[\textit{n.b.} this is similar to the compliant wall boundary conditions imposed in linear stability analyses, see][]{Carpenter1990}.  This assumption is not rigorously justified and breaks down if the compliant wall significantly alters the near-wall mean flow.  As mentioned previously, in a more complete model, the mean flow must be consistent with the total Reynolds stress generated by the modes to enable \textit{a priori} predictions of changes in the mean profile.  This would allow for changes in the Reynolds stress contributions from the modes to be propagated through the system via an iterative process.  Unfortunately, the convergence properties of this iterative process are not known and as such are the subject of ongoing research.

The mechanical admittance $Y$ dictates the relative phase and amplitude of the wall-normal velocity and the pressure at the wall, which can be estimated based on the structural properties of the wall.  As an example, the dynamics of the simple spring- and damper-backed membrane shown in Fig.~\ref{fig1:schematic} can be expressed as:

\begin{equation}\label{eqWallMotion}
-p_\kb(0) = (-C_m \omega^2 -i\om C_d + C_k)\eta_\kb
\end{equation}

{\noindent}where

\begin{equation}\label{eqWallParameters}
C_m = \frac{\rho_w}{\rho h},\: C_d = \frac{d_w}{\rho u_\tau}, \: C_k = \frac{k_w h}{\rho u_\tau^2},
\end{equation}

{\noindent}are the dimensionless mass, damping, and spring coefficients respectively; $\rho$ is the mass density of the fluid, $\rho_w$ is the mass of the wall per unit area, $k_w$ is the effective spring constant, and $d_w$ is the damping ratio.  From (\ref{eqWallMotion}), it is easy to show that the mechanical admittance for this wall is:

\begin{equation}\label{eqAdmittance}
Y = \frac{v_\kb(0)}{p_\kb(0)} = \frac{i\om}{-C_m \omega^2 -i\om C_d + C_k} = \frac{i\om (C_k - \om^2 C_m)-\om^2 C_d}{(C_k-\om^2 C_m)^2+\om^2 C_d^2}.
\end{equation}

{\noindent}While (\ref{eqWallMotion}) assumes that the membrane is of negligible stiffness and tension-free, both of these effects can be introduced via terms involving spatial derivatives.  Specifically, the restoring effects of tension ($C_t$) and stiffness ($C_s$) would introduce additional terms proportional to $(\kx^2+\kz^2)$ and $(\kx^4+2\kx^2\kz^2+\kz^4)$, respectively.  In essence, this would result in a wavenumber-dependent effective spring constant $C_{ke} = C_k + C_t(\kx^2+\kz^2) + C_s(\kx^4+2\kx^2\kz^2+\kz^4)$.  Similarly, more complex walls involving viscoelastic coatings can be incorporated by solving a wave equation for the solid layer with appropriate boundary conditions \citep[\eg][]{GadelHak2000}.

Thus, the admittance $Y$ can be used to account for the effects of compliant walls with known material properties.  However, instead of following the traditional approach involving a trial-and-error procedure with different wall properties, the present study considers the inverse problem: finding an optimal wall that leads to the most favorable effect on the turbulent flow structures that are known to be important in real flows.  The concept of a complex wall admittance has been used previously \citep{Landahl1962}, perhaps most notably by \citet{Sen1988}, who employed a similar inverse approach to study the stability of laminar boundary layer flows over compliant surfaces.

In this paper, a simple gradient-free pattern search is used to find the optimal $Y$ \citep{Hooke1961}, with a particular focus on resolvent modes resembling the NW cycle and VLSMs.  Optimality in this context is defined in two different ways: (i) the compliant wall that leads to the greatest reduction in forcing-response gain ($\sk$) relative to the rigid-wall case, indicating mode suppression, and (ii) the compliant wall that leads to the greatest reduction in the channel-integrated Reynolds stress contribution from the resolvent mode:

\begin{equation}\label{eqReynoldsStress}
RS = \int\limits_{0}^{2} \sk^2 \mathrm{re}(u_\kb^* v_\kb) (y-1)\,dy,
\end{equation}

{\noindent}where $\mathrm{re}()$ denotes the real component.  The weighing factor $(y-1)$ is included per the decomposition developed by \citet{Fukagata2002}, who showed that the friction coefficient in turbulent channel flow comprises a laminar component inversely proportional to the Reynolds number, defined using the bulk-averaged velocity, and a turbulent component proportional to the weighted Reynolds stress integral shown in (\ref{eqReynoldsStress}).  For channels with compliant walls, the drag coefficient also comprises an additional term which depends on the wall deformation.  However, the integrated Reynolds stress still plays a major role \citep{Nakanishi2012}.  Note that (\ref{eqReynoldsStress}) represents a forcing-normalized quantity; it assumes unit forcing along the first forcing mode, such that the velocity response at that wavenumber-frequency combination is given by $\sk \uk(y)$.

Keep in mind that the above optimization of wall properties is only valid for a single wavenumber-frequency combination.  As discussed in \S \ref{sec:results-Previous}, the optimal admittance identified for one mode could very well have adverse effects on resolvent modes at other wavenumber-frequency combinations.  Further, the admittance of the same wall also changes across spectral space.  For instance, the simple second-order model (\ref{eqAdmittance}) shows that $Y=f(\om)$.  Accounting for the effects of tension and stiffness would also introduce wavenumber dependencies, $Y = f(\kx,\kz)$.

\subsection{Numerical implementation}\label{sec:model-Validation}
The discretized resolvent operator (\ref{eqForcingResponse}) is constructed using a spectral collocation method on Chebyshev points.  The differentiation matrices are computed using the MATLAB differentiation matrix suite developed by \citet{Weideman2000}.  The mean velocity profile in the resolvent operator $U(y)$ is obtained based on the turbulent viscosity model proposed by \citet{Reynolds1967}:

\begin{equation}\label{eqEddyViscosity}
\nu_T = 
\frac{1}{2} \left[ 1 + \left( \frac{\kappa \Ret}{3}(2y-y^2)(3-4y+2y^2) \left( 1-e^{(|y-1|-1)\frac{\Ret}{\alpha}} \right) \right) \right]^{1/2}-\frac{1}{2},
\end{equation}

{\noindent}where $\nu_T$ is normalized by $\nu$.  The parameter $\kappa$ is the von Karman constant, and $\alpha$ appears in van Driest's mixing length equation.  The values for these parameters are taken to be $\kappa = 0.426$ and $\alpha = 25.4$, based on the optimization performed by \citet{Moarref2012} to fit the mean profile obtained from channel flow DNS at $\Ret = 2000$ \citep{Hoyas2006}.  The majority of the results presented in this paper correspond to this Reynolds number.

Unlike in a pipe, the SVD of the channel flow resolvent operator generally yields pairs of structurally similar response modes with near-identical singular values but differing symmetry along the channel centerline \citep{Moarref2013}.  To avoid this pairing and to make the computation more efficient, the grid is restricted to $N$ Chebyshev points in the lower half-channel with user-specified symmetry across the centerline.  All of the results presented in this paper correspond to resolvent modes with antisymmetric $u_\kb$, $w_\kb$, $p_\kb$ fields and symmetric $v_\kb$.  The differentiation matrices for the velocity and pressure fields are modified to account for these symmetry conditions \citep[see][]{Trefethen2000}.

A grid resolution study showed that the singular values converge to within $O(10^{-4})$ for $N\ge 100$, which is the minimum resolution employed in this study.  For $N \ge 100$, the singular values for symmetric and antisymmetric modes were also found to differ by less than $O(10^{-4})$ at the wavenumber-frequency combinations considered in this paper.  Similarly, the gradient-free pattern search procedure to find the optimal $Y$ has a specified tolerance of $O(10^{-4})$.  As a rough estimate of computational expense, each resolvent evaluation (\ie, setting up the operator with the appropriate boundary conditions and performing the SVD) takes approximately $0.1 s$ on one core of a laptop for $N=100$ and $0.5 s$ for $N=200$.

\section{Results and discussion}\label{sec:results}
While the resolvent formulation can be used to evaluate the effect of compliant walls on modes across spectral space, this section initially focuses on the two most natural starting points: the NW cycle (\S\ref{sec:results-NW}) and VLSMs (\S\ref{sec:results-VLSM}).  This is followed by discussions concerning the effects of mode speed (\S\ref{sec:results-Speed}), previous DNS results and the overall efficacy of the optimal walls identified in \S\ref{sec:results-NW}-\ref{sec:results-VLSM} (\S\ref{sec:results-Previous}), and more realistic wall properties (\S\ref{sec:results-Walls}).

\subsection{Near-wall modes}\label{sec:results-NW}
The dynamically important NW cycle is characterized by coherent structures with streamwise length scale $\lxp \approx 1000$, spanwise length scale $\lzp \approx 100$, and characteristic propagation speed $c^+ \approx 10$ \citep{Robinson1991}.  As a result, this section considers the design of compliant walls that interact favorably with the resolvent mode corresponding to $\kb = (\kx,\kz,c^+) = (12,\pm 120,10)$ at $\Ret = 2000$.  The wavenumbers are normalized with respect to the channel half-height, such that the streamwise and spanwise wavelengths for this mode are $\lxp = 2\pi\Ret/\kx \approx 1050$ and  $\lzp = 2\pi\Ret/\kz \approx 105$, respectively.  Note that $\Ret = 2000$ was chosen because it represents the upper limit for which DNS results are readily available \citep{Hoyas2006}, and the lowest Reynolds number at which the VLSMs considered in the following section become prominent \citep{Smits2011}.  Resolvent modes resembling the NW cycle at the lower Reynolds numbers typical of DNS studies with compliant walls (\eg $\kb \approx (1,10,10)$ at $\Ret = 140 - 180$) exhibit very similar behavior to that described below.

\begin{figure}
	\centering
	\includegraphics[width=12cm]{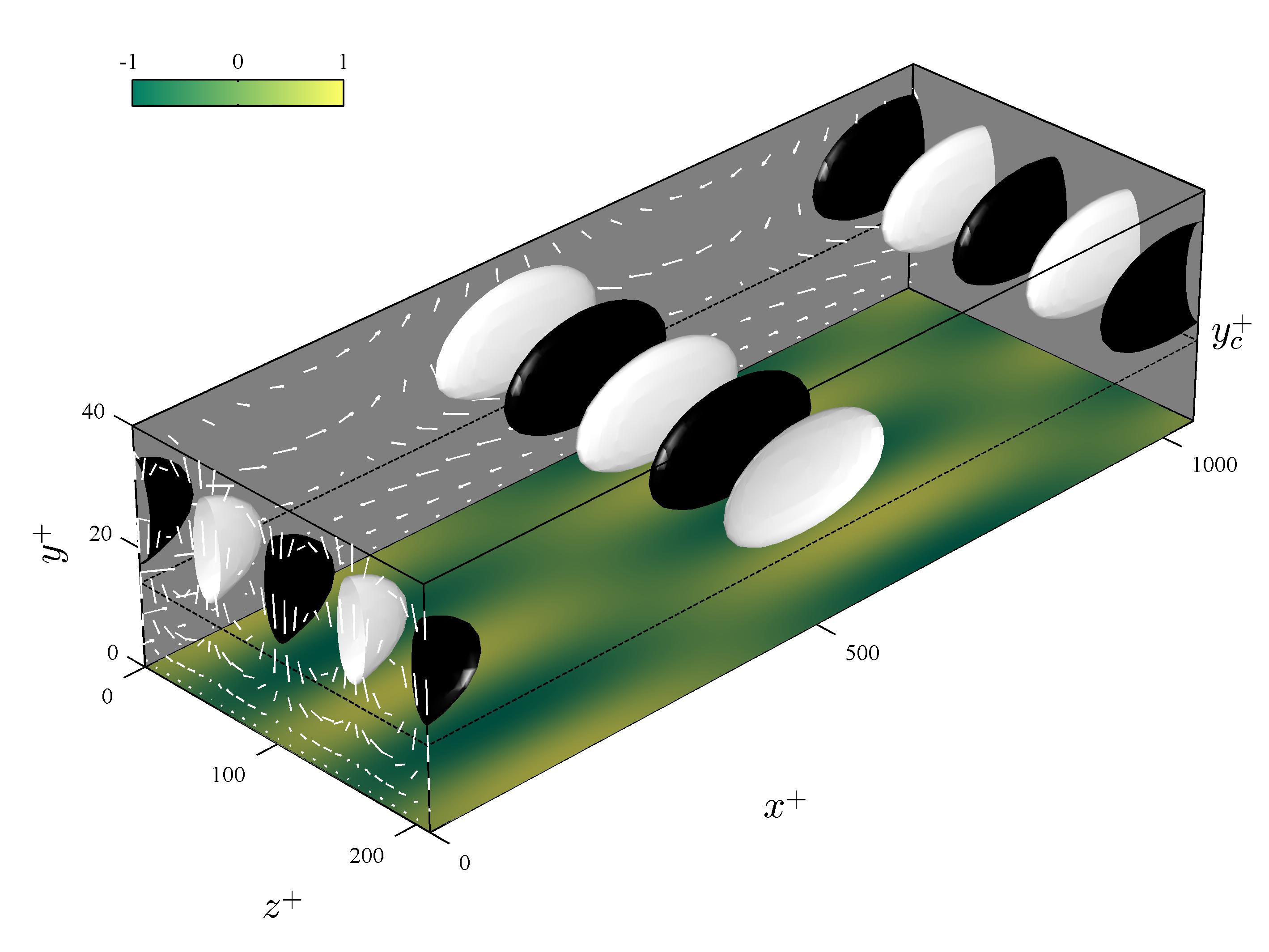}
	\caption{Velocity structure for the NW-type mode $\kb = (\kx,\kz,c^+) = (12,\pm 120,10)$ at $\Ret = 2000$ over a rigid wall.  The black and white isosurfaces show negative and positive wall-normal velocities at 80\% of maximum absolute value.  The shading at the wall represents the normalized wall-pressure field.  Note that the velocity structure shown in this figure includes contributions from both the $\kz=120$ and $\kz=-120$ modes.}
	\label{fig2:NW-Null}
\end{figure}

The null-case velocity field associated with these NW-type resolvent modes is shown in Fig.~\ref{fig2:NW-Null}.  As expected for Fourier modes, the velocity field shows regions of alternating positive and negative velocity with length scales $\lxp$ and $\lzp$.  Consistent with known features of the NW cycle, the velocity field in the spanwise-wall normal plane (see $x^+ = 0$) shows counter-rotating quasi-streamwise vortices localized just above the critical layer $\ypc \approx 15$, where the mode speed matches the local mean velocity, $U(\ypc)=c^+$.  The velocity field in the streamwise-wall normal plane (see $z^+ = 0$) exhibits alternating prograde (with the mean shear) and retrograde rotation.  The wall-pressure field lags the wall-normal velocity field by $\approx \pi/2$, such that regions of maximum (minimum) wall-pressure coincide with regions of increasing (decreasing) wall-normal velocity.

\begin{figure}
	\centering
	\includegraphics[width=12cm]{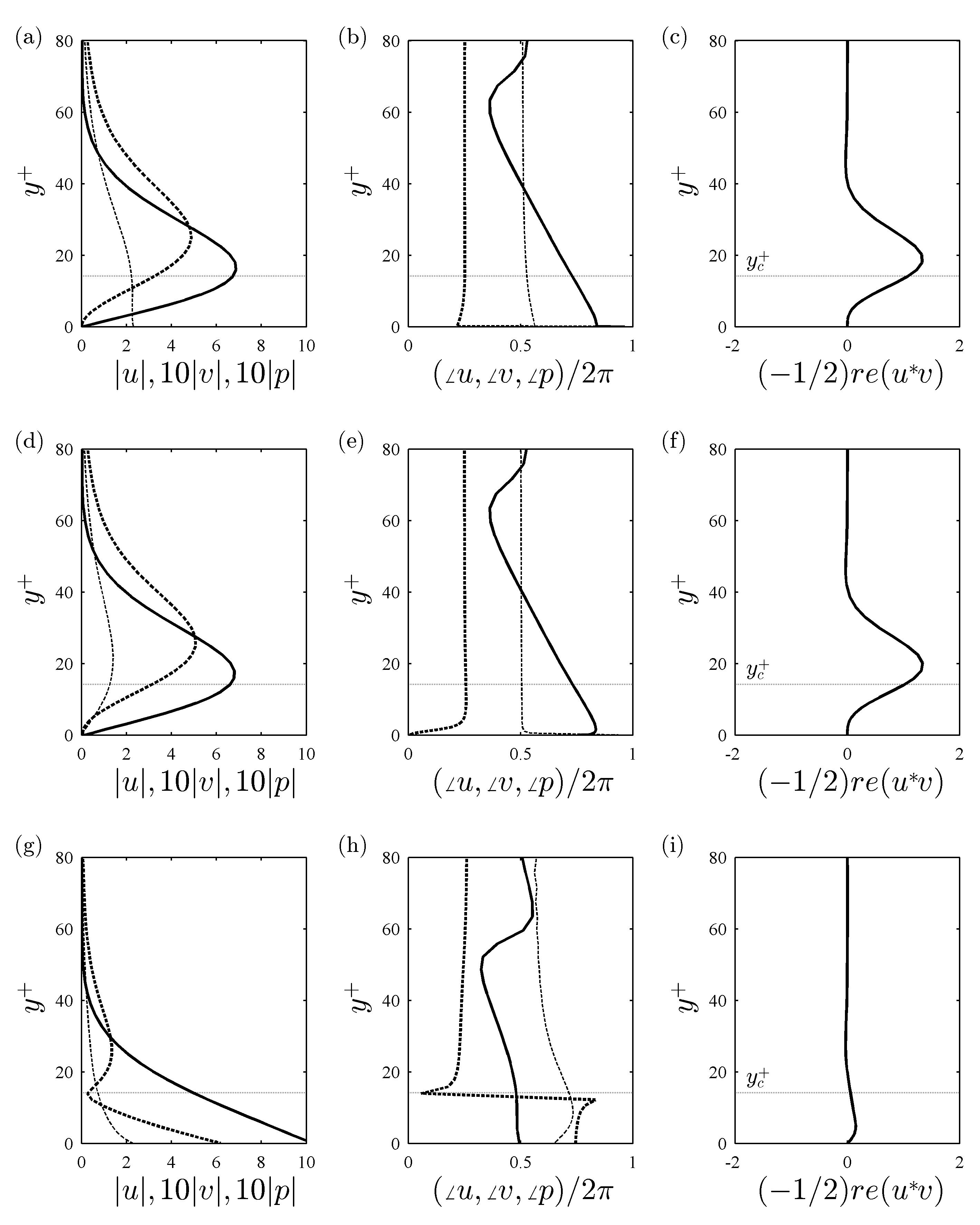}
	\caption{Profiles showing the wall-normal variation in amplitude (a,d,g) and phase (b,e,h) for the streamwise velocity (solid lines), wall-normal velocity (dashed lines) and pressure fields (fine dashed lines) for the resolvent modes resembling the NW cycle.  The normalized Reynolds stress contribution is shown in plots (c,f,i).  Plots (a-c) represent the rigid wall case, (d-f) represent the optimal wall in terms of singular value suppression ($Y = 1.92+0.55i$), and (g-i) represent the optimal wall in terms of Reynolds stress reduction ($Y = 2.23+1.55i$).}
	\label{fig3:NW-Lines}
\end{figure}

The amplitude, phase, and Reynolds stress profiles shown in Fig.~\ref{fig3:NW-Lines}(a-c) provide further insight into the structure of this mode.  The streamwise velocity field peaks at the critical layer while the wall-normal velocity peaks at a location slightly farther from the wall.  The streamwise and wall-normal velocity are $\approx \pi$ out of phase at the critical layer, which means that the Reynolds stress contribution from this mode (Fig.~\ref{fig3:NW-Lines}c) also peaks near $\ypc$.  As discussed in \citet{McKeon2010}, these features are typical of high-gain resolvent modes.

Note that the wall-normal velocity and pressure fields exhibit a near-constant $\pi/2$ phase difference.  \citet{Luhar2014b} show that this is because the pressure field associated with the resolvent modes arises primarily from the so-called \textit{fast} source terms in the Poisson equation, which represent the linear interaction between the mean shear and wall-normal velocity, \ie $\nabla_\kb^2 p_\kb \approx -2i\kx v_\kb (\p U/\p y)$.  It can be shown that the \textit{slow} source term in the pressure Poisson equation corresponds to the divergence of the nonlinear forcing, $\nabla_\kb \cdot \fk$, which tends to be near-zero for the resolvent modes.  In other words, the most amplified forcing fields tend to be divergence-free.  However, despite the fact that the resolvent pressure fields do not contain contributions from the nonlinear slow terms, they reproduce many previous observations \citep[\eg local pressure minima under hairpin heads, pressure maxima associated with regions where sweeps meet ejections etc.;][]{Luhar2014b}.

The pattern-search optimization procedure indicates that a compliant wall with admittance $Y = 1.92 + 0.55i$ leads to the greatest reduction in singular value for this mode.  Specifically, this wall yields a $32\%$ reduction in gain (\ie the ratio of compliant- to rigid-wall singular values is $\sigma_{\kb c}/\sigma_{\kb 0} = 0.68$).  As shown in Fig.~\ref{fig3:NW-Lines}d-f, this reduction in singular value is not accompanied by a significant change in mode structure.  The only major distinction is that the magnitude of the pressure field at the wall is much lower over the compliant wall (Fig.~\ref{fig3:NW-Lines}d, fine dashed line).  Perhaps this reduction in the magnitude of the wall-pressure field indicates a link to the vorticity-flux control method proposed by \citet{Koumoutsakos1999}, who showed that a suppression of the spanwise vorticity flux from the wall led to a significant (up to $40\%$) drag decrease in channel flow DNS at $\Ret = 180$.  Since the planar gradients of the wall-pressure field ($i\kx p_\kb(0)$ and $i\kz p_\kb(0)$) are directly proportional to the flux of streamwise and spanwise vorticity from the wall, the reduction in wall-pressure suggests a reduction in the vorticity flux.  Although, keep in mind that a direct comparison between the present results and the vorticity flux control method is not strictly appropriate.  The controller developed by \citet{Koumoutsakos1999} only considered the flux of spanwise vorticity and so the resulting flow structures were more similar to the spanwise-constant traveling waves discussed in \S \ref{sec:results-Previous}.  Finally, note that the normalized Reynolds stress contribution from this mode does not change significantly relative to the rigid-wall case (Fig.~\ref{fig3:NW-Lines}c,f).  However, for identical forcing strengths, the $32\%$ suppression in mode amplitude over the compliant wall would translate into a near-$50\%$ reduction in the actual Reynolds stress generated by this mode ($\propto \sk^2$, see \ref{eqReynoldsStress}).

\begin{figure}
	\centering
	\includegraphics[width=12cm]{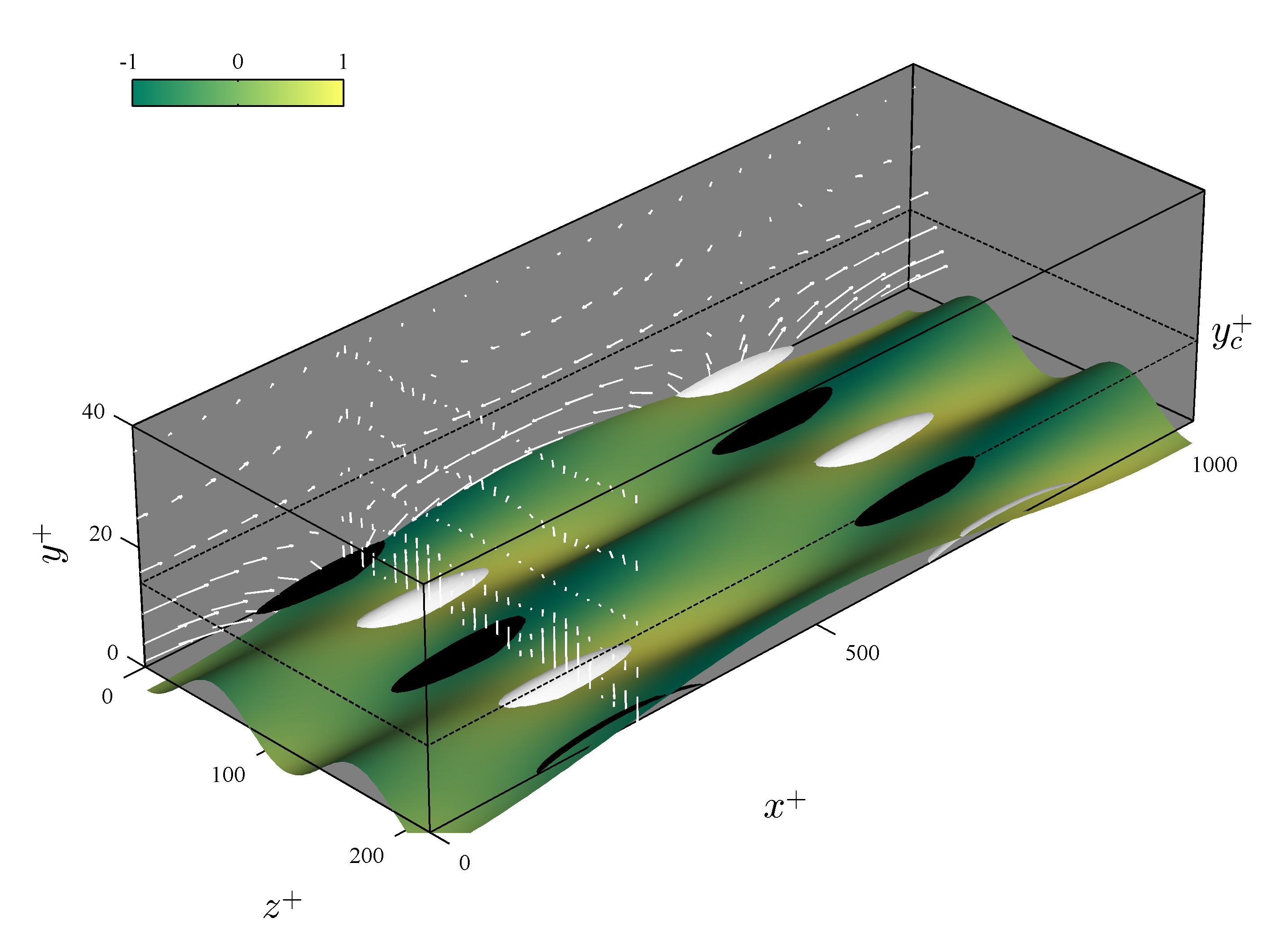}
	\caption{Velocity structure for the NW-type mode with the optimal Reynolds stress-reducing wall.  The black and white isosurfaces show negative and positive wall-normal velocities at 80\% of maximum absolute value.  The shading on the compliant wall (deflection not to scale) represents the normalized pressure field.}
	\label{fig4:NW-Drag}
\end{figure}

A compliant wall with admittance $Y = 2.23 + 1.55i$ is found to be optimal based on the lowest channel-integrated Reynolds stress criterion (\ref{eqReynoldsStress}).  Although this wall does not yield a substantial reduction in gain ($\sigma_{\kb c}/\sigma_{\kb 0} = 0.90$), it does result in a significant modification of the flow structure (Fig.~\ref{fig3:NW-Lines}g-i).  Instead of peaking at or near the critical layer, the magnitude of the velocity fields is largest at the wall.  Further, the phase relationship between the streamwise and wall-normal velocity is modified substantially relative to the rigid-wall case (Fig.~\ref{fig3:NW-Lines}h).  The velocity fields are approximately $\pm \pi/2$ out of phase across $\ypc \approx 0-40$, leading to a significant reduction in the normalized Reynolds stress (Fig.~\ref{fig3:NW-Lines}i).  Physically, this change in mode structure means that the quasi-streamwise vortices observed for rigid-wall case, which are important for Reynolds stress and TKE production via the lift-up mechanism, are almost entirely suppressed and structures resembling rollers are generated near the wall (Fig.~\ref{fig4:NW-Drag}).

Recall that the above results were obtained with linear approximations to the boundary conditions (\ref{eqKinematicBCu}-\ref{eqKinematicBCw}).  Since the velocity magnitudes peak at the wall (Fig.~\ref{fig3:NW-Lines}), the boundary conditions clearly play a major role in dictating mode behavior.  Hence, the use of linearized boundary conditions is a significant limitation of the present approach, which needs to be carefully evaluated against results from simulations employing more accurate boundary conditions (\eg involving immersed boundaries or coordinate transformations) or from experiments as they become available.

\begin{figure}
	\centering
	\includegraphics[width=12cm]{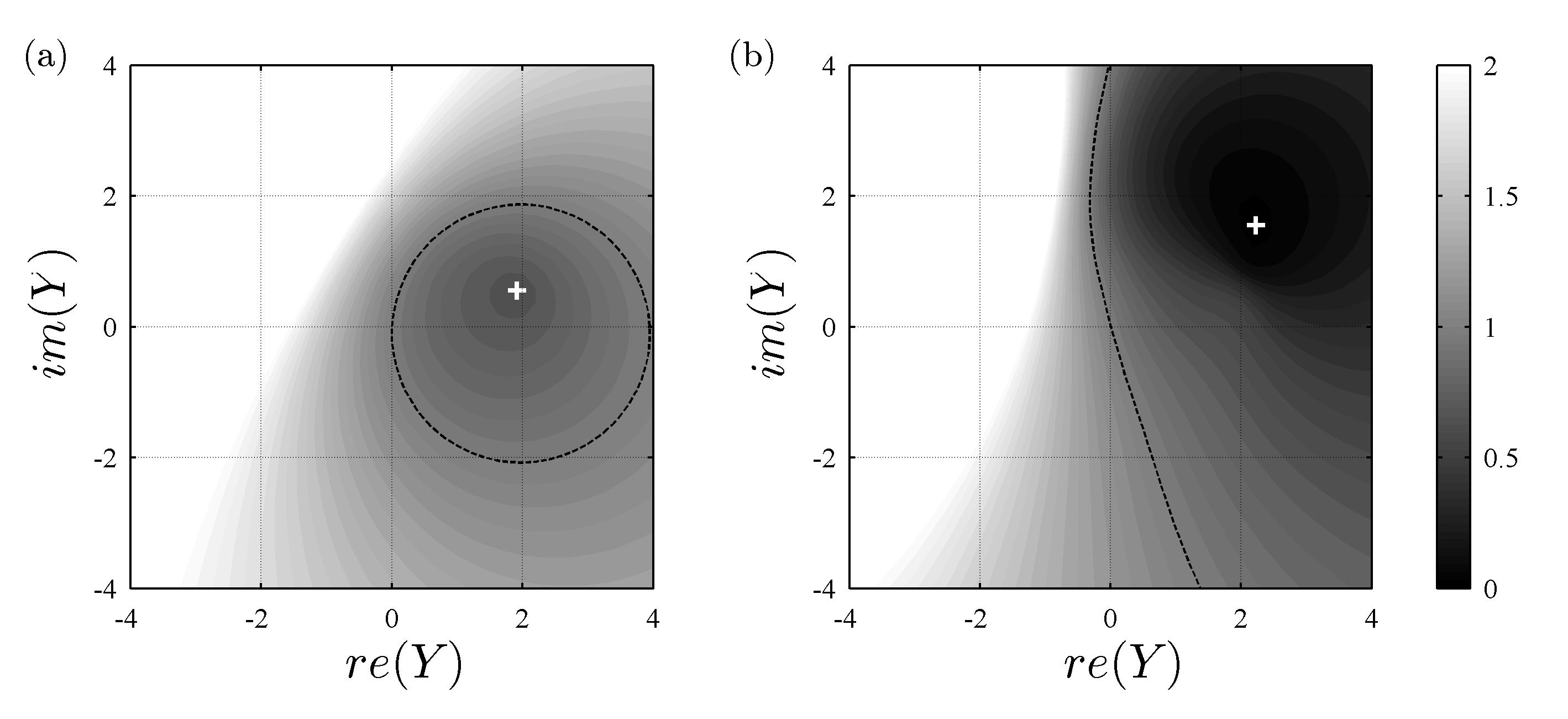}
	\caption{Contours showing the ratio of compliant- to rigid-wall singular values $\sigma_{\kb c}/\sigma_{\kb 0}$ (a) and channel-integrated Reynolds stress $RS_{\kb c}/RS_{\kb 0}$ (b) across a range of wall admittances, $Y$, for resolvent modes resembling the NW cycle. The optimal admittances identified based on the pattern search procedure are marked with a white $+$.  The channel-integrated Reynolds stress is defined in (\ref{eqReynoldsStress}).  The dashed contour line corresponds to a ratio of 1, \ie no change.}
	\label{fig5:NW-minYmap}
\end{figure}

The results presented thus far suggest that it might be possible to suppress the dynamically-important NW cycle using compliant walls.  However, it is important to the keep in mind that the admittance $Y$ must reflect the physical properties of the wall.  The simple spring-damper model (\ref{eqAdmittance}) suggests that the optimal admittances identified in this section, $Y = 1.92 + 0.55i$ for mode suppression and $Y = 2.23 + 1.55i$ for Reynolds stress reduction, would require negative damping coefficients, \ie $\mathrm{re}(Y)>0$ corresponds to $C_d<0$.  This is further confirmed by the contour maps shown in Fig.~\ref{fig5:NW-minYmap}.  Only compliant walls with $\mathrm{re}(Y)>0$ lead a reduction in singular value (Fig.~\ref{fig5:NW-minYmap}a).  A reduction in Reynolds stress does appear possible with $\mathrm{re}(Y)<0$, despite the increase in gain (see $Y \approx -0.1+2i$, Fig.~\ref{fig5:NW-minYmap}b).  However, this result must be treated with caution because the higher-rank resolvent modes not considered in this paper may become important as the singular values increase relative to the rigid-wall case, and the Reynolds stress contribution from these higher-rank modes could offset the Reynolds stress-reduction from the rank-1 modes.  In other words, the rank-1 approximation might fail in the modified flow.

Since the NW cycle is a dominant feature of low Reynolds number flows, the results presented in this section may also explain why previous DNS studies have met with limited success.  The present analysis suggests that the positive-damping walls typically tested in DNS are unlikely to interact favorably with the NW cycle (to our knowledge, no previous simulations have considered negative damping coefficients).  Optimal wall admittances with $\mathrm{re}(Y)>0$ are also indicative of a situation where the wall-normal velocity is, at least partially, in-phase with the wall-pressure field (Fig.~\ref{fig4:NW-Drag}).  This phase relationship is consistent with suggestions made in previous studies based on heuristic arguments \citep[\eg][]{Xu2003,Fukagata2008} that the velocity fluctuations need to be in phase with the wall-pressure for drag reduction.

Finally, it is also important to keep in mind that a wall with negative damping is, strictly speaking, no longer passive.  It results in a net energy transfer into the flow.  However, negative damping does not necessarily imply that the wall requires an external energy input.  It may be possible to identify materials that are passive globally but exhibit negative damping over a limited bandwidth.

\subsection{Very-large-scale motions}\label{sec:results-VLSM}

\begin{figure}
	\centering
	\includegraphics[width=12cm]{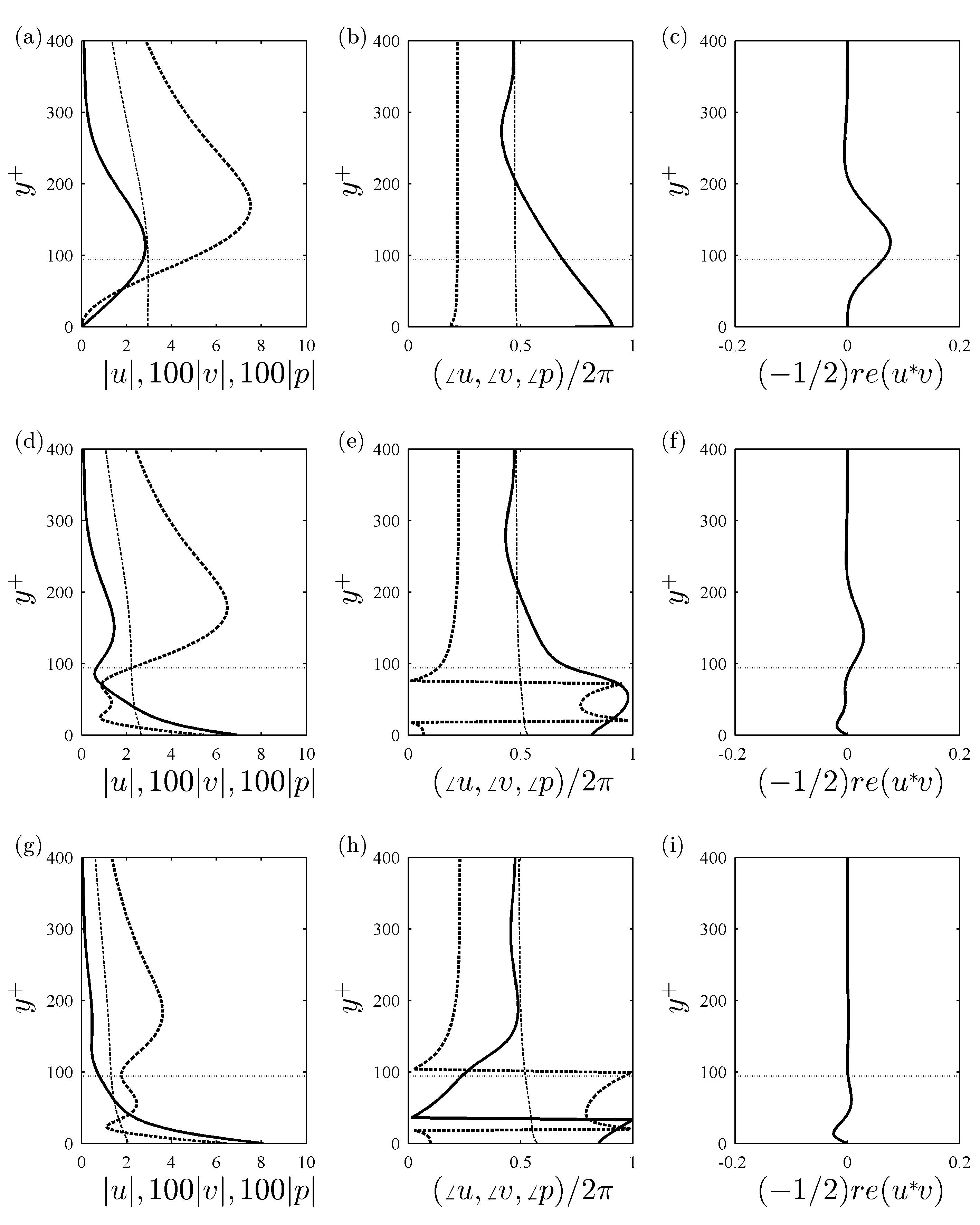}
	\caption{Profiles showing the wall-normal variation in amplitude (a,d,g) and phase (b,e,h) for the streamwise velocity (solid lines), wall-normal velocity (dashed lines) and pressure fields (fine dashed lines) for the resolvent modes resembling VLSMs.  The normalized Reynolds stress contribution is shown in plots (c,f,i).  Plots (a-c) represent the rigid wall case, (d-f) represent the optimal wall in terms of singular value suppression ($Y = -2.04-0.44i$), and (g-i) represent the optimal wall in terms of Reynolds stress reduction ($Y = -3.10-0.39i$).}
	\label{fig6:VLSM-Lines}
\end{figure}

\begin{figure}
	\centering
	\includegraphics[width=12cm]{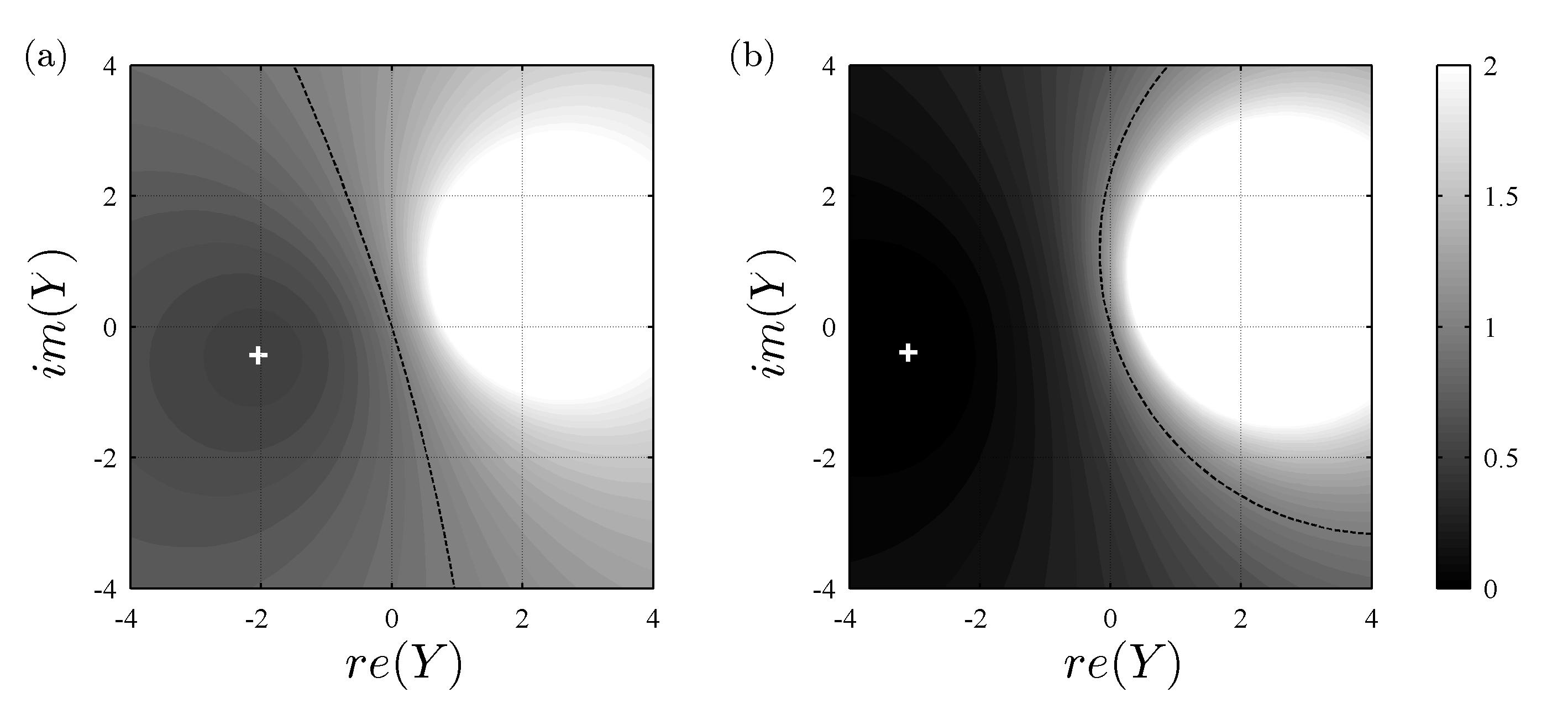}
	\caption{Contours showing the ratio of compliant- to rigid-wall singular values $\sigma_c/\sigma_0$ (a) and channel-integrated Reynolds stress $RS_c/RS_0$ (b) across a range of wall admittances, $Y$, for resolvent modes resembling VLSMs. The optimal admittances identified based on the pattern search procedure are marked with a white $+$.  The dashed contour line corresponds to a ratio of 1, \ie no change.}
	\label{fig7:VLSM-minYmap}
\end{figure}

This section considers the effect of compliant surfaces on modes with $\kb = (1,\pm 10,16)$ at $\Ret = 2000$.  Similar to the VLSMs \citep{Monty2009a,Smits2011}, these modes represent flow structures of streamwise and spanwise length scales $\lambda_x \approx 6h$  ($\lxp \approx 12600$) and $\lambda_z \approx 0.6h$ ($\lzp\approx 1260$), propagating downstream at $c^+ = 16$.  Despite the roughly ten-fold increase in wavelength, Fig.~\ref{fig6:VLSM-Lines}a-c show that the rigid-wall structure for this resolvent mode is very similar to that for the NW-mode considered in the previous section.  The streamwise velocity, wall-normal velocity, and Reynolds stress again peak at or near the critical layer, which is now located at $\ypc \approx 94$, and there is a near-constant $\pi/2$ phase difference between $v_\kb$ and $p_\kb$.

However, unlike the modes resembling the NW cycle, compliant walls with positive damping interact favorably with these larger-scale modes.  This is illustrated by the predicted performance maps shown in Fig.~\ref{fig7:VLSM-minYmap}; regions of singular value and Reynolds stress reduction are primarily confined to admittances with $\mathrm{re}(Y)<0$.  The pattern search suggests that the optimal wall admittances for gain and Reynolds stress reductions for this mode are $Y = -2.04-0.44i$ and $Y = -3.19-0.39i$, respectively.  Both admittances lead to similar reductions in singular value and changes in mode structure (see Fig.~\ref{fig6:VLSM-Lines}d-f and g-i).  The best gain-reducing wall leads to $\skc/\sko = 0.52$, while the optimal Reynolds stress reducing wall leads to $\skc/\sko = 0.56$.

The mode structure over the compliant walls is modified such that the streamwise velocity (solid lines) is largest at the wall, while the magnitude of the wall-normal velocity (bold dashed lines) exhibits local maxima above and below the critical layer.  The local peak in $v_\kb$ below the critical layer is also accompanied by a $\pm \pi/2$ phase shift between the wall and the critical layer (see bold dashed lines, Fig.~\ref{fig6:VLSM-Lines}e,h).  Note that the normalized Reynolds stress contribution from the compliant-wall resolvent modes is significantly lower compared to the rigid-wall case.  Fig.~\ref{fig6:VLSM-Lines}i suggests that the compliant wall may even lead to a reversal in the sign of the Reynolds stress.

\begin{figure}
	\centering
	\includegraphics[width=12cm]{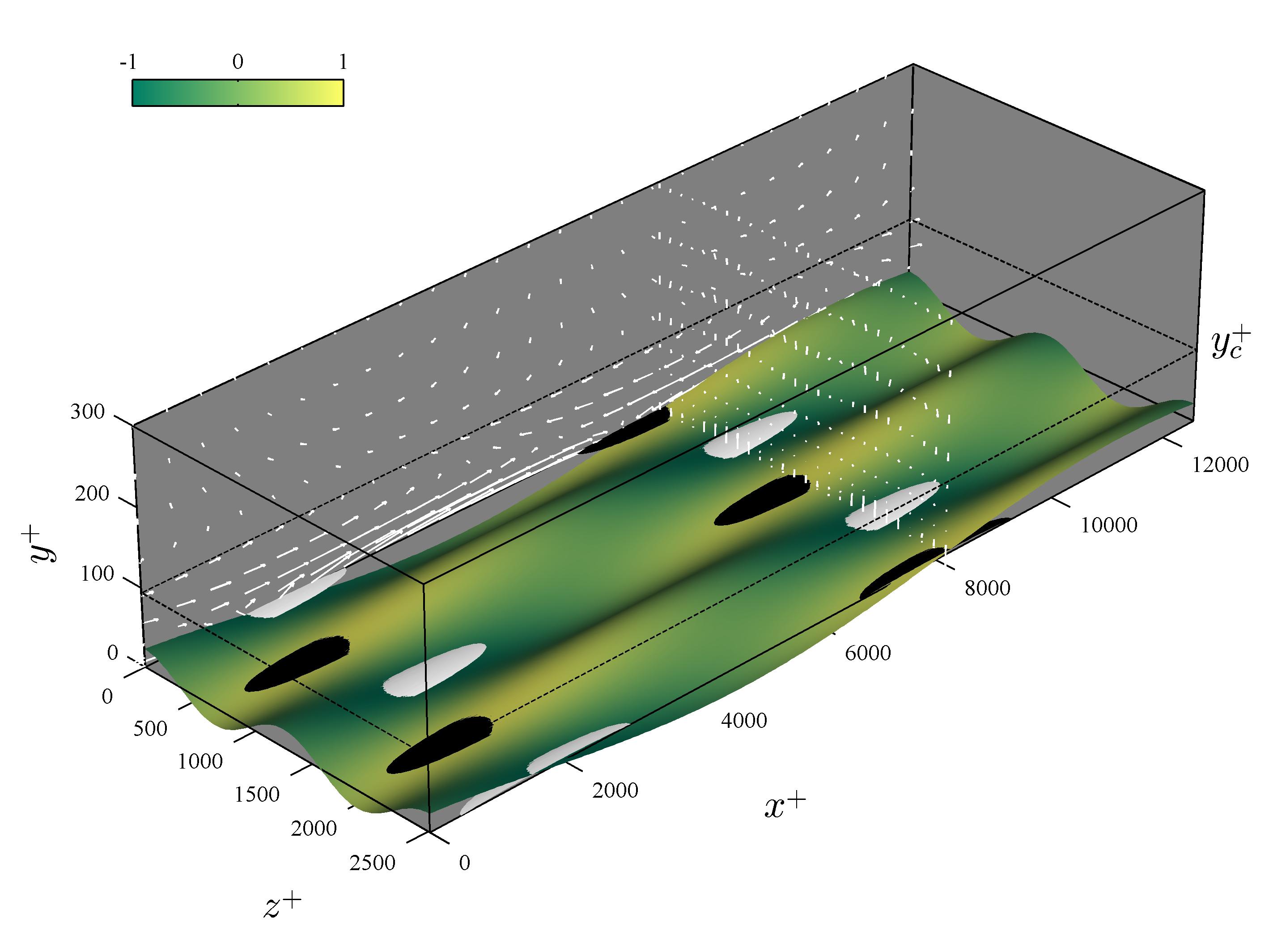}
	\caption{Velocity structure for the VLSM-type mode with the optimal Reynolds stress-reducing wall.  The black and white isosurfaces show negative and positive wall-normal velocities at 80\% of maximum absolute value.  The shading on the compliant wall (deflection not to scale) represents the normalized pressure field.}
	\label{fig8:VLSM-Drag}
\end{figure}

The rigid-wall flow structure associated with this large-scale mode is broadly similar to that shown in Fig.~\ref{fig2:NW-Null}, but with vastly different length scales.  Fig.~\ref{fig8:VLSM-Drag} shows how the wall optimized for Reynolds stress alters this physical structure.  The compliant wall again leads to a suppression of the quasi-streamwise vortices associated with Reynolds stress production.  However, unlike the NW-modes, the wall-normal velocity and pressure fields are not in phase at the wall.  Regions of positive $p_\kb$ coincide with regions of negative $v_\kb$ (\ie the wall moves downwards under high pressure), as expected for $\mathrm{re}(Y)<0$.

The fact that walls with positive damping are predicted to interact favorably with the larger VLSM-type modes might explain why previous experiments have met with more success than numerical simulations.  In general, the experiments \citep{Lee1993,Choi1997} have been carried out at higher Reynolds numbers, where these larger-scale flow structures play a more prominent role.  Since the VLSMs are known to have an organizing influence on the near-wall turbulence \citep{Marusic2010}, they may also serve as a pathway for compliant walls to influence the entire flow.

\subsection{Effect of mode speed}\label{sec:results-Speed}
As a sensitivity analysis, this section evaluates the effects of compliant walls on resolvent modes with varying speed $c^+$.  The speed $c^+$ determines the wall-normal localization of the mode \citep{McKeon2010}.  Slower-moving resolvent modes are located close to the wall and so they have stronger source terms in the pressure Poisson equation (recall that the fast source term is $\propto v_\kb \p U/\p y$ and the mean velocity gradient is larger close to the wall).  \citet{Luhar2014b} show that, everything else being equal, the stronger source terms and wall-proximity translate into a larger wall-pressure signature for slower-moving resolvent modes.  As the mode speed increases, the flow structure associated with the resolvent modes moves further away from the wall and the magnitude of the wall-pressure field decreases.  Above some threshold speed, the modes detach from the wall entirely and have a near-zero wall-pressure field.  This threshold speed tends to be wavelength-dependent; longer modes have a larger wall-normal extent, which means that they detach from the wall at higher mode speeds.

\begin{figure}
	\centering
	\includegraphics[width=8cm]{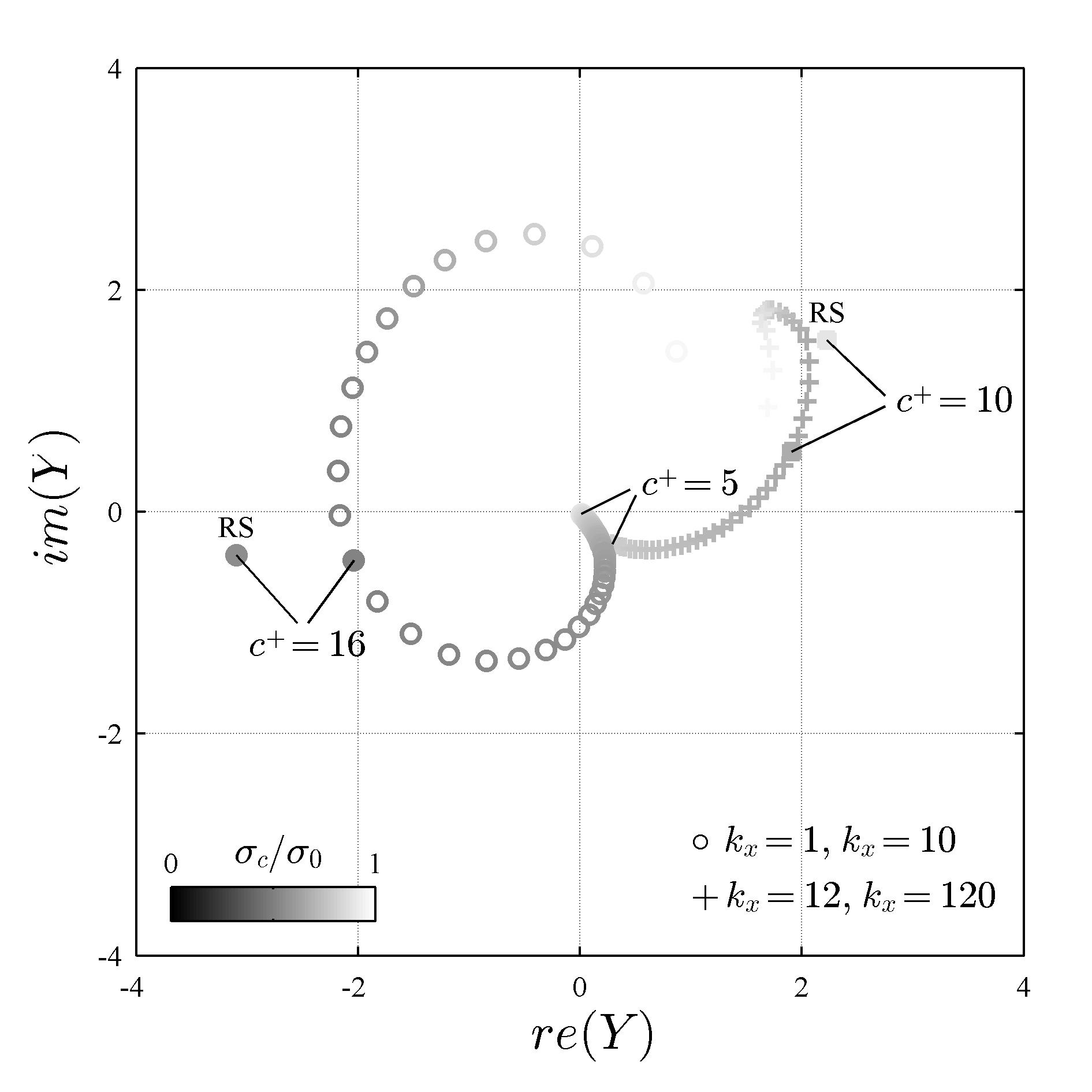}
	\caption{Optimal wall admittances, in terms of singular-value suppression, for modes of varying speed $c^+ = 5-20$.  Each marker represents an increment of $\Delta c^+ = 0.2$.  The color of the marker indicates the degree of suppression.  Also shown are the optimal admittances for Reynolds stress reduction for the NW and VLSM-type modes considered in \S\ref{sec:results-NW} and \S\ref{sec:results-VLSM} (marked with an RS).}
	\label{fig9:minY-Speed}
\end{figure}

The above observations suggest that relatively rigid walls with low $|Y|$ are likely to interact only with slower-moving modes with high fluctuating wall-pressure, while softer walls might be able to interact with modes further away from the wall.  However, it may be impossible to influence detached modes with a compliant wall.  This is confirmed by the data shown in Fig.~\ref{fig9:minY-Speed}, which plots the optimal admittance for gain-reduction for modes with the same wavelengths as the NW- and VLSM-type structures considered in the previous two sections, $(\kx, \kz) = (12,120)$ and $(\kx, \kz) = (1,10)$, but mode speed varying from $c^+ = 5$ to $c^+ = 20$.  The magnitude of the optimal admittance, $|Y|$, is smallest for the slowest- modes and increases with increasing mode speed.  In other words, stiffer walls are optimal for slow-moving modes with high amplitude wall-pressure fields while softer walls are better suited for faster modes further away from wall.  Note that there is a turning point near $c^+ \approx 19$ for the longer $\kx = 1$ (Fig.~\ref{fig9:minY-Speed}, $\circ$) modes and $c^+ \approx 13$ for the $\kx = 12$ modes (Fig.~\ref{fig9:minY-Speed}, $+$), above which the optimal $|Y|$ decreases and the effectiveness of the compliant surfaces diminishes (\ie the markers get lighter).  Consistent with the mode detachment hypothesis, it may not be possible to design walls that can interact with these fast-moving structures.

Figure~\ref{fig9:minY-Speed} also provides a measure of the sensitivity of the optimal wall admittances on mode parameters.  The fact that there are no sharp jumps in the optimal $Y$ with increasing $c^+$ indicates that the optimal wall admittances are unlikely to be very different for resolvent modes similar to those identified in the previous two sections.  This is important because the previous two sections assume that the NW cycle and VLSMs are adequately characterized by resolvent modes at one specific wavenumber-frequency combination.  However, in reality the NW cycle and the VLSMs represent \textit{regions} that are energetic in spectral space.  Hence, resolvent modes with slightly different wavelengths and speeds are equally viable candidates to serve as models for the NW cycle and VLSMs.

Finally, note that there is a consistent trend in the optimal pressure-velocity phase relationship, $\angle Y = \angle v_\kb(0) - \angle p_\kb(0)$, with mode speed.  Perhaps the most interesting aspect of this trend is that the optimal admittance for the larger $(\kx,\kz)=(1,10)$ modes is also characterized by $\mathrm{re}(Y)>0$ (\ie $C_d < 0$, see \ref{eqAdmittance}) for $c^+ \lesssim 14$.  It may be the case that the negative-damping requirement discussed in \S\ref{sec:results-NW} is restricted to slower-moving modes.  These issues will be explored further in future publications.

\subsection{Comparison with previous DNS and overall efficacy of optimal walls}\label{sec:results-Previous}
So far, this paper has focused primarily on the interaction between compliant walls and individual resolvent modes which resemble flow structures that are energetic in natural wall turbulence.  However, it is important to keep in mind that the compliant wall could also have detrimental effects elsewhere in spectral space.  This is illustrated by the recent DNS results obtained by \citet{Kim2014} at $\Ret = 140$.  The softest compliant wall tested by \citet[][termed case II]{Kim2014} led to the generation of large-amplitude quasi-2D waves of wavelength $\lambda_x \approx 2.4h$ propagating downstream at a speed roughly $0.3\times$ the centerline velocity ($c^+ \approx 5$).  These waves, which are clearly not energetic in rigid-wall turbulent channel flow, led to a substantial increase in the total drag.  \Citet{Kim2014} attributed the generation of the quasi-2D waves to a wall-resonance effect.

\begin{figure}
	\centering
	\includegraphics[width=12cm]{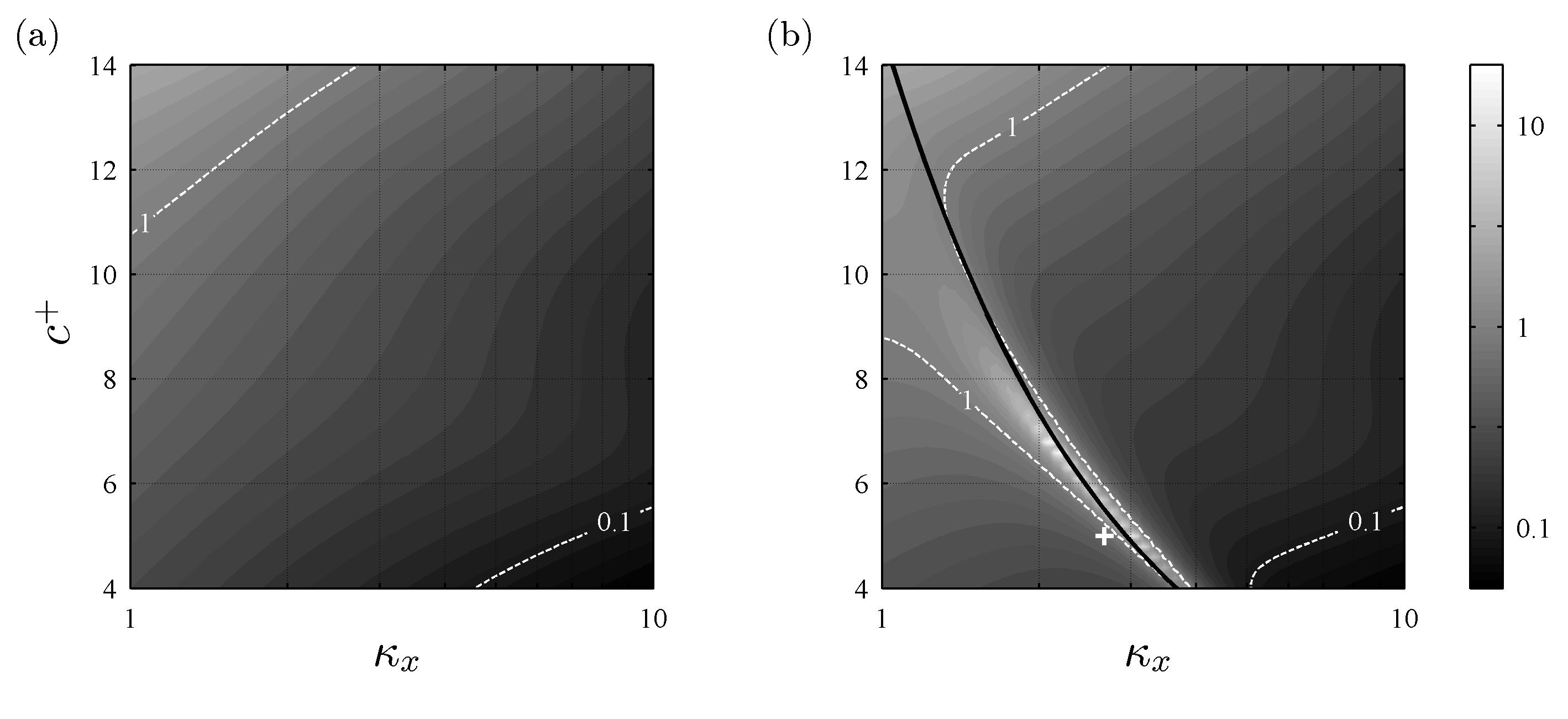}
	\caption{Singular values for spanwise-constant ($\kz = 0$) modes as a function of streamwise wavenumber and wave speed at $\Ret = 140$.  (a) Shows the rigid-wall singular values  and (b) shows the singular values for the case II compliant wall tested by \citet{Kim2014}.  The $+$ marker represents the length and velocity scale for the quasi-2D traveling wave observed in the DNS.  The solid black line shows the resonant frequency of the wall $\om_r$.  Note the log-scaling for the shading.}
	\label{fig10:Kim2014}
\end{figure}

Figure~\ref{fig10:Kim2014} shows that the framework developed in this study can predict this resonant response with minimal computation.  Fig.~\ref{fig10:Kim2014}a shows the rigid-wall singular values for 2D ($\kz = 0$) resolvent modes at $\Ret = 140$ as a function of the streamwise wavenumber and wave-speed.  Fig.~\ref{fig10:Kim2014}b shows the singular values predicted by the present analysis over the case II compliant wall of \citet{Kim2014}.  Since \citet{Kim2014} also employed a simple mass-spring-damper mechanical model for the compliant wall, the admittances were estimated using (\ref{eqAdmittance}) to arrive at these predictions.  The mass coefficient for the case II compliant wall was $C_m = 2$, while the spring- and damping-coefficients, normalized by $1.5\times$ the bulk-velocity ($\approx 21 u_\tau$), were $C_k^* = 1$ and $C_d^* = 0.5$, respectively.  This translates into $C_k \approx 440$ and $C_d \approx 10.5$ with the $u_\tau$ normalization shown in (\ref{eqWallParameters}).  The resonant frequency for this wall is $\om_r = \om_n \sqrt{1-2\zeta^2}= 14.7$.  Here, $\om_n = \sqrt{C_k/C_m}$ is the undamped natural frequency of the wall and $\zeta = C_d/(2\sqrt{C_k C_m})$ is the damping factor.

The compliant-wall singular values clearly show a region of high-amplification around $\om_r$ extending from $(\kx,c^+) \approx (3,5)$ to $(\kx, c^+) \approx (1.5,9)$ which does not exist for the rigid-wall case.  The singular values over the compliant wall are as much as 70 times larger relative to the rigid case.  The greatest change in singular value relative to the rigid-wall case falls at $\kx = 2.8$ and $c^+ = 5.25$, which corresponds closely to the wavelength and speed ($\kx = 8/3$ and $c^+=4.78$) of the quasi-2D traveling wave observed in DNS.  The slight differences in wavenumber and speed could be attributed to three different reasons: (i) the uncertainty in $C_k$ and $C_d$ arising from the bulk-velocity to friction velocity translation of these dimensionless coefficients, (ii) the use of the unperturbed mean velocity profile in the resolvent operator, and (iii) the linearized boundary conditions at the wall.  The latter two sources of uncertainly are significant because the DNS results show a substantial increase in drag over the compliant wall, which is likely to be accompanied by a change in the mean velocity profile \citep[not reported by][]{Kim2014}, and the 2D waves excited had amplitudes in excess of $10$ viscous units.

\begin{figure}
	\centering
	\includegraphics[width=12cm]{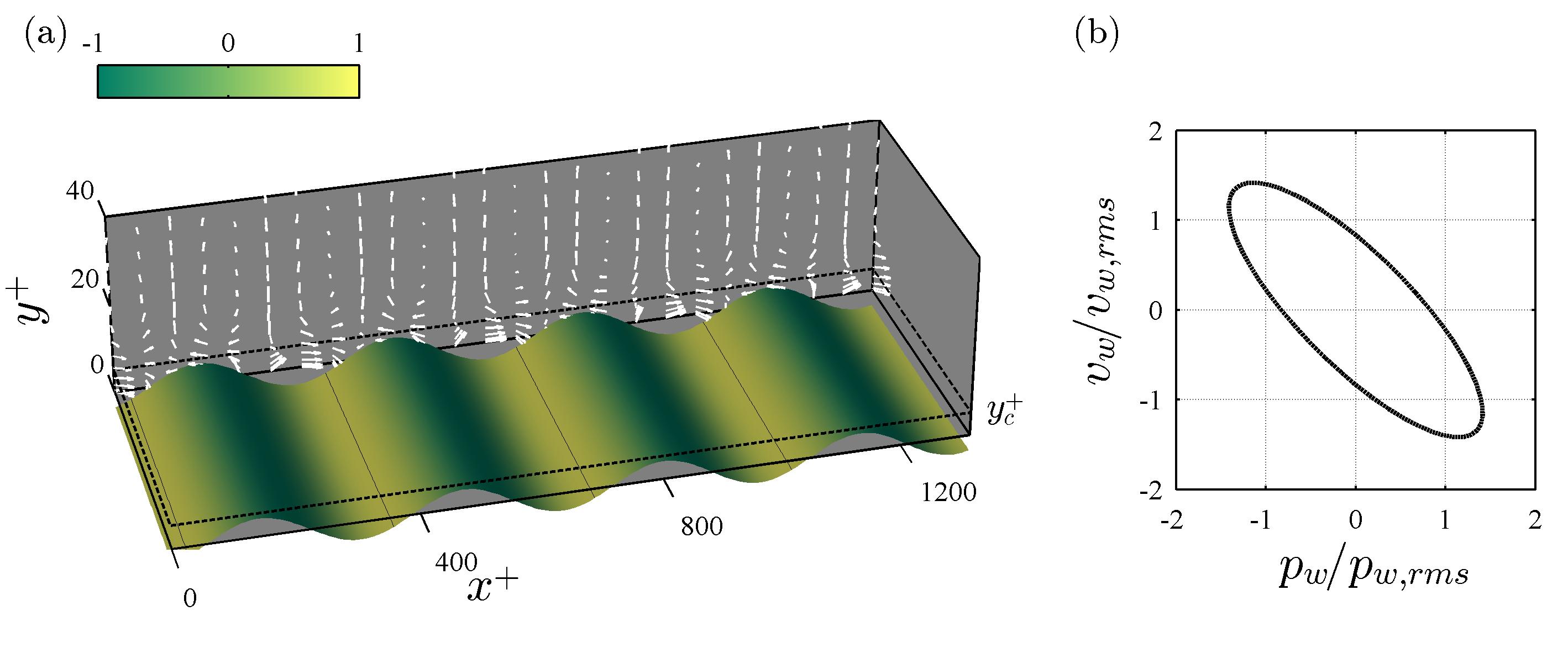}
	\caption{(a) Velocity structure for the resolvent mode corresponding to wavenumber-frequency combination $(\kx,\kz,c^+) = (8/3,0,5)$ at $\Ret = 140$.  This represents roughly the quasi-2D traveling waves observed in DNS by \citet{Kim2014}.  The shading on the compliant wall (deflection not to scale) represents the normalized pressure field.  (b) The phase relationship between wall-normal velocity and pressure at the wall.}
	\label{fig11:Kim2014-Structure}
\end{figure}

Figure~\ref{fig11:Kim2014-Structure} shows the flow structure and $v-p$ relationship associated with this resonant resolvent mode.  The streamwise velocity for this mode is confined to a very small region close to the wall ($y^+<10$).  Above this location, the velocity field comprises a predominantly up-down motion.  This is consistent with the DNS results, which showed that the streamwise velocity field at $y^+ = 14$ above the compliant wall was dominated by streaky structures elongated in the streamwise direction \citep[Fig. 8 in][]{Kim2014}.  Further, the predicted phase relationship between the wall-normal velocity and pressure at the wall (Fig.~\ref{fig11:Kim2014-Structure}b) is also similar to the DNS observations \citep[Fig. 7 in][]{Kim2014}.

Although qualitative in nature, the comparison presented above indicates that the framework developed in this paper is a useful first-order test of material properties before implementation in more computationally intensive numerical simulations or experiments.  Recall that each resolvent evaluation only takes $\approx 0.1s$ on one core of a laptop.  The contour maps shown in Fig.~\ref{fig10:Kim2014} were estimated at 100 linearly spaced intervals in $c^+$ and 50 log-spaced intervals in $\kx$, and as such required approximately $2\times$ $500 s$ of computation time on one core of a laptop (\textit{n.b.} without any significant effort towards making the computation efficient).  Of course, the above computations were limited by the constraint $\kz = 0$.  A true \textit{a priori} evaluation requires a sweep across the entire spectral space.  However, this is still inexpensive relative to numerical simulations, especially so given the spectral sparsity observed recently in wall turbulence \citep{Bourguignon2014}.

\begin{figure}
	\centering
	\includegraphics[width=12cm]{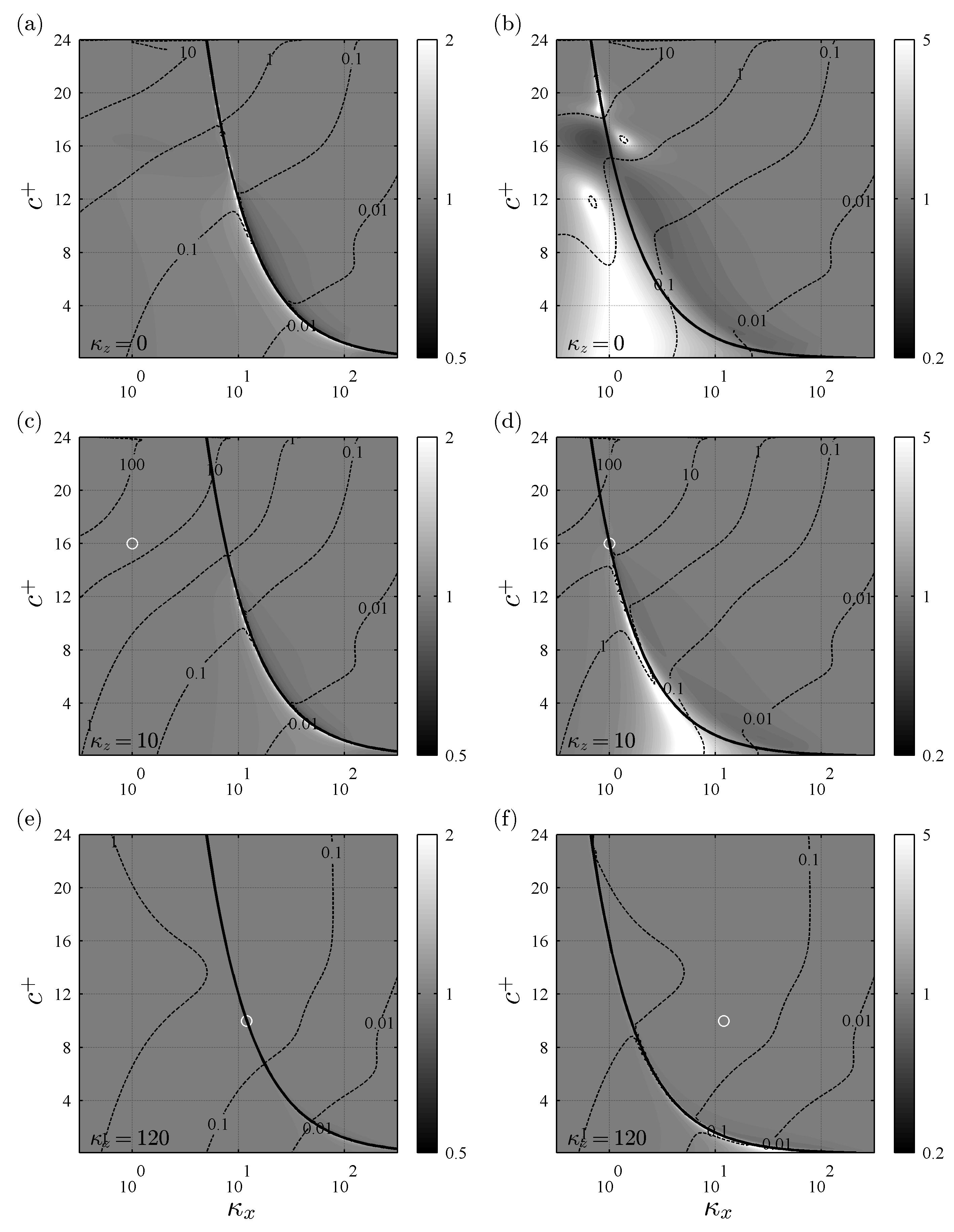}
	\caption{Shaded contours showing the singular value ratio $\skc/\sko$ as a function of wavenumber $\kx$ and speed $c^+$  at $\kz = 0$ (spanwise constant; a,b), $\kz = 10$ ($\lzp \approx 1250$; c,d) and $\kz = 120$ ($\lzp \approx 100$; e,f).  White regions denote an increase in mode amplification over the compliant wall.  Darker regions denote suppression.  Note the log-scaling for the shading.  The dashed contour lines indicate the magnitude of the singular values above the compliant wall, $\skc$.  Panels (a,c,e) correspond to the compliant wall optimized for modes resembling the NW cycle ($\kx=12$, $\kz = 120$, $c^+ = 10$; white $\circ$ in e,f).  Panels (b,d,f) correspond to the optimal compliant wall for VLSM-type modes ($\kx=1$, $\kz = 10$, $c^+ = 16$; white $\circ$ in c,d). The solid black lines show the resonant frequency of the walls $\om_r$.}
	\label{fig12:Re2000-OptimalWalls-Broadband}
\end{figure}

The high-amplitude resonant modes observed in DNS suggest that compliant walls optimized to suppress modes resembling the NW cycle and the VLSMs could lead to detrimental effects at other wavenumber-frequency combinations.  If the increased drag contribution from these other modes outweighs the decrease from the NW- and VLSM-type modes, the compliant wall would have a negative net influence on performance.  To test whether this is the case, the effect of the optimal gain-reducing walls identified in \S\ref{sec:results-NW} and \S\ref{sec:results-VLSM} on modes elsewhere in spectral space is considered next.

Assuming a dimensionless mass coefficient of $C_m = 2$, it can be shown based on (\ref{eqAdmittance}) that the optimal wall for the NW-type modes ($Y = 1.92+0.55i$) requires a high stiffness coefficient $C_k = 28817$ and a negative damping coefficient $C_d = -0.48$.  Similarly, the optimal wall for the VLSM-type modes ($Y=-2.04-0.44i$) requires $C_k = 510$ and $C_d = 0.47$.  Note that the resonant frequencies of the optimal walls correspond very closely to the temporal frequencies for the modes resembling the NW cycle ($\om_r = 120.03$, $\om = 120$) and the VLSMs ($\om_r = 15.97$, $\om = 16$).  This is to be expected given that the amplitude of the compliant wall response is likely to be largest (\ie largest $|Y|$) at or near the resonant frequency (\ref{eqAdmittance}).  This large wall response would translate into a bigger impact on the turbulent flow structure.

Figure~\ref{fig12:Re2000-OptimalWalls-Broadband} shows how the optimal walls affect the amplification of resolvent modes with $\kz = 0$, $\kz = 10$ and $\kz = 120$ as a function of streamwise wavenumber and mode speed at $\Ret = 2000$.  As expected, the compliant walls have the largest effect, positive or negative, on modes with frequencies close to $\om_r$.  Further, the impact of the compliant wall is restricted to a smaller region of spectral space around $\om_r$ as the spanwise wavelength decreases ($\kz$ increases, compare Fig.~\ref{fig12:Re2000-OptimalWalls-Broadband}a,b with Fig.~\ref{fig12:Re2000-OptimalWalls-Broadband}e,f) and the mode speed $c^+$ increases (compare $c^+\approx 4$ with $c^+ \approx 12$ in Fig.~\ref{fig12:Re2000-OptimalWalls-Broadband}d).  This is because, everything else being equal, larger-scale resolvent modes that are slow-moving  and localized close to the wall have a much larger wall-pressure signature compared to smaller, faster modes farther from the wall \citep{Luhar2014b}.  The higher amplitude wall-pressure field yields a larger wall response, which compensates for the decrease in $|Y|$ as the temporal frequency of the modes moves away from $\om_r$.  This scale- and speed-dependence may also explain why the compliant wall optimized for the NW-type modes (Fig.~\ref{fig12:Re2000-OptimalWalls-Broadband}a,c,e) has a smaller reach in spectral space compared to the wall-optimized for the VLSM-type modes (Fig.~\ref{fig12:Re2000-OptimalWalls-Broadband}b,d,f).  The higher resonant frequency for the wall optimized for NW-type modes means that it interacts preferentially with modes that are faster or shorter in $x$ ($\om = c^+ \kx \approx \om_r$).  These modes are likely to have a smaller wall-pressure signature and so only interact significantly with the compliant wall at conditions close to resonance.

Note that there are sharp transitions in the effectiveness of the compliant walls around the resonant frequency. In general modes with $\om < \om_r$ are further amplified by the compliant wall while modes with $\om > \om_r$ are suppressed (although, this appears to reverse for $c^+ \gtrsim 15$).  Thus, modes with length and velocity scales very similar to the assumed parameters for NW cycle or VLSMs may actually be further amplified over the optimal compliant walls.  This result suggests that it may be better for the overall performance to design compliant walls that are slightly detuned and resonant away from the spectral region of interest.  More generally, the above results suggest that it is insufficient to optimize compliant wall-properties for a single wavenumber-frequency combination; the optimization must be performed over a wider region in spectral space.  This is similar to the approach employed in designing compliant walls for transition delay, wherein compliant wall properties are optimized against all the mode types which the fluid and wall can concurrently support \citep[see \eg ][]{Dixon1994}.

\subsection{More realistic walls and surface instabilities}\label{sec:results-Walls}
\begin{figure}
	\centering
	\includegraphics[width=12cm]{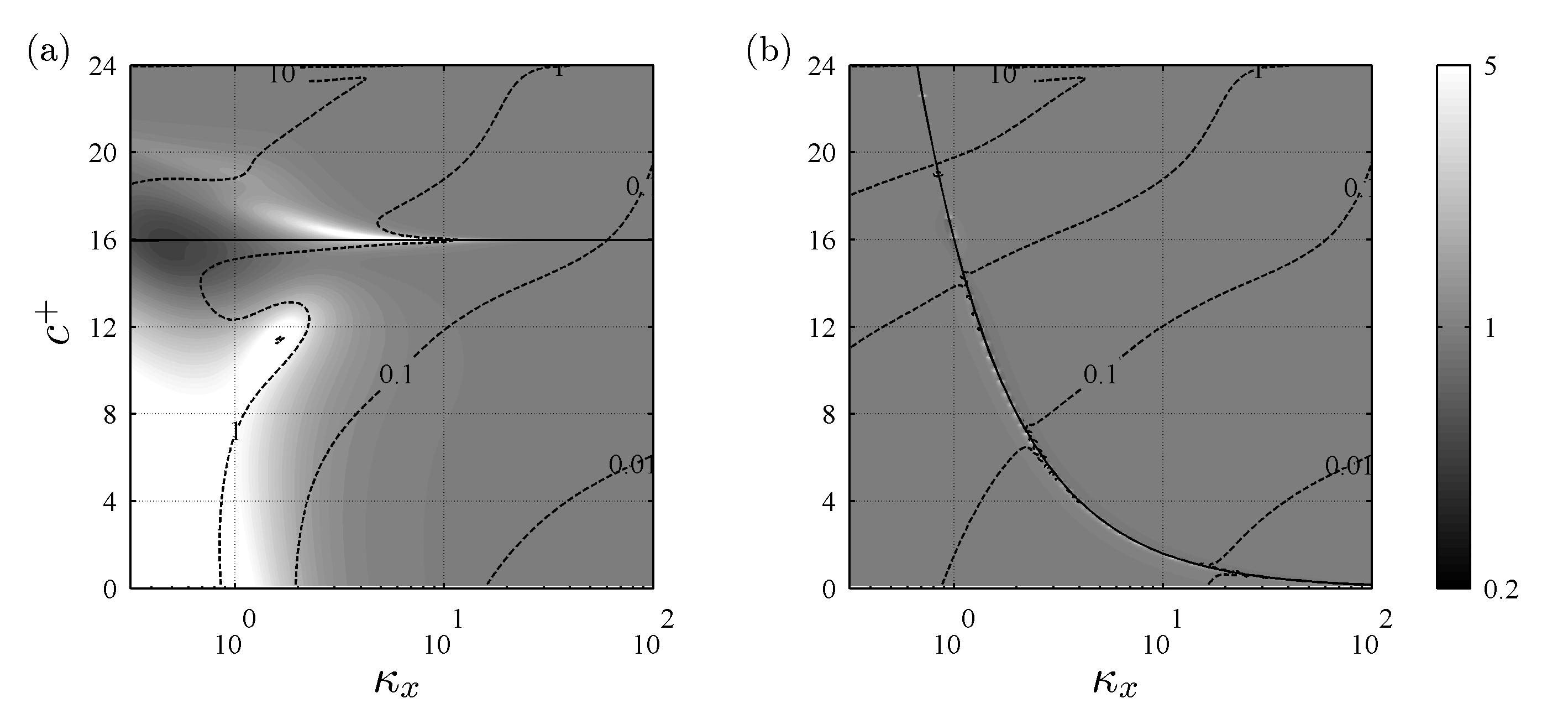}
	\caption{Shaded contours showing the singular value ratio $\skc/\sko$ as a function of wavenumber $\kx$ and speed $c^+$  for spanwise constant $\kz = 0$ modes.  Both panels correspond to walls optimized to suppress the VLSMs (compare with Fig.~\ref{fig12:Re2000-OptimalWalls-Broadband}b).  Panel (a) represents a compliant wall in streamwise tension $C_{tx}$ with spring constant $C_k = 0$ and (b) corresponds to a mass-spring-damper wall with mass ratio $C_m = 200$.  White regions denote an increase in mode amplification over the compliant wall.  Darker regions denote suppression.  Note the log-scaling for the shading.  The dashed contour lines indicate the magnitude of the singular values above the compliant wall, $\skc$.  The solid black lines show the free wave speed, $c^+_w$, for the tensioned wall in (a) and the resonant frequency $\om_r$ in (b).}
	\label{fig13:Re2000-ComplexWalls}
\end{figure}

Much of the discussion thus far has been based on the simple mass-spring-damper wall model.  As illustrated by Fig.~\ref{fig12:Re2000-OptimalWalls-Broadband}, this results in fluid--structure interactions that are primarily dependent on frequency.  In addition, these dynamics are local; the wall does not communicate in the streamwise or spanwise directions.  This means that the wall cannot support wave propagation and is non-dispersive.  Practically constructed walls such as the elastic and viscoelastic surfaces tested in experiment not only support waves but also have an embedded natural length scale (\eg layer thickness).  This allows for spatio-temporal matching between the fluid and solid motions, leading to much stronger interactions.  The experiments performed by Gad-el-Hak and co-workers \citep{GadelHak1984,GadelHak1986} show that such elastic and viscoelastic layers support two different kinds of surface instabilities under laminar and turbulent boundary layers: static divergence and traveling wave flutter.  The so-called static divergence waves are very slow moving ($<5\%$ of free stream velocity) with long wavelengths and large amplitudes, while traveling wave flutter tends to have shorter wavelengths and lower amplitudes but propagates much faster, at speeds comparable to the free shear-wave speed in the solid.  Figure~\ref{fig13:Re2000-ComplexWalls}a shows that the resolvent framework qualitatively predicts similar effects.

Specifically, Fig.~\ref{fig13:Re2000-ComplexWalls}a shows the effect of a compliant wall optimized to suppress VLSMs ($Y=-2.04-0.44i$ at $\kx=1$, $\kz=10$ and $c^+ = 16$) on spanwise constant modes, similar to the results shown in Fig.~\ref{fig12:Re2000-OptimalWalls-Broadband}b.  However, the wall is assumed to be in streamwise tension $C_{tx}=510$ rather than on a spring support.  As a result, the effective spring constant for the wall is $C_{ke}=C_{tx} \kx^2$ and the admittance is $Y=i\om (C_{tx}\kx^2-C_m \om^2 - i C_d)^{-1}$.  Note that this wall can support streamwise propagating waves.  Relative to the rigid-wall case, there are two regions of extremely high amplification over the compliant wall: for very long slow-moving modes ($\kx \lesssim 1$ and $c^+ \lesssim 8$) and for shorter modes ($\kx = 2-10$) at speeds close to the undamped wave speed for the wall $c^+_w \approx \sqrt{C_{tx}/C_m} = 16$.  It can be argued that the first class of modes are similar to static divergence waves and the second class of resonant modes are similar to traveling wave flutter \citep[see also][]{Sen1988,Riley1988}.  Keep in mind that the tensioned wall does not have a natural length scale and so the wavelengths of the high-amplification regions are likely determined by the channel height.

For more quantitative comparisons with experiments, the resolvent formulation needs to be extended to account for boundary conditions that more accurately reflect elastic and viscoelastic coatings.  This is the subject of ongoing work.  Even so, direct comparisons with the observed high amplitude static divergence waves may be difficult given that the framework developed in this paper is limited to linearized boundary conditions.

Finally, note that the preceding results are based on an assumed mass coefficient $C_m = 2$.  While $C_m \sim O(1)$ is appropriate for liquid flows over compliant walls, $C_m \sim (10^3)$ for gas flows over walls with realistic material properties.  Compliant walls are unlikely to be effective for systems with such large mass ratios because the magnitude of the admittance decreases sharply with increasing $C_m$ (\ref{eqAdmittance}) away from resonance, resulting in much smaller wall responses to the fluid pressure perturbations.  This is illustrated by the results shown in Fig.~\ref{fig13:Re2000-ComplexWalls}b.  Again, this figure shows the effect of a wall optimized to suppress the VLSMs on spanwise constant modes.  However, the mass ratio here is $C_m = 200$.  Unlike the results shown in Fig.~\ref{fig12:Re2000-OptimalWalls-Broadband}b, this wall has a minimal effect on resolvent modes with frequencies far from $\om_r$.  In other words, the wall is effectively rigid at this high $C_m$ limit, suggesting that the fluid and solid impedances must at least match approximately for the compliant surface to have a significant effect on the flow.

\section{Conclusions and outlook}\label{sec:conclusion}
In many ways, compliant surfaces are a tremendously attractive proposition for the passive control of wall-bounded turbulent flows.  Such walls would require no complex sensing or actuation machinery and, depending on the coating materials required, may even be relatively inexpensive.  However, progress towards designing an effective compliant surface has been hindered by the lack of a computationally inexpensive theoretical framework that can adequately characterize the interaction between the turbulent flow structures and the wall.  The results presented in this study suggest that the resolvent framework developed by McKeon, Sharma and co-authors \citep{McKeon2010,Sharma2013,Moarref2013,Luhar2014a,Luhar2014b} offers a potential solution.

Unlike previous attempts at low-order models, the resolvent framework stems directly from the governing NSE.  It does not make any ad-hoc assumptions that are hard to justify physically and there are no Reynolds number restrictions beyond the requirement of a mean velocity profile.  However, it is also important to keep in mind some of its limitations.  Chief among these is the fact that the nonlinear convective terms in the NSE are simply treated as unstructured forcing in the present paper.  A complete low order model would require an explicit treatment of the nonlinear coupling across resolvent modes.  In addition, this paper only considers the effect of compliant walls on the gain and structure of modes expected to be energetic in turbulent flows.  A reduction in gain would translate into mode suppression and a reduction in turbulent kinetic energy, which is a useful indicator of performance.  However, this approach cannot predict the eventual effect on the mean velocity profile or skin friction without further assumptions.  Another weakness is the use of the linearized kinematic boundary conditions (\ref{eqKinematicBCu}-\ref{eqKinematicBCw}), which lose accuracy with increasing wall deformation.  Despite these simplifications, \S\ref{sec:results-Previous} shows that the resolvent framework is able to predict the resonant quasi-2D response observed in recent compliant wall turbulent channel flow DNS \citep{Kim2014}.  Therefore, at the very least, this framework can serve as a first-order test of material properties before implementation in more computationally-expensive simulations or physical experiments.

At the same time, the results presented in \S\ref{sec:results-NW}-\S\ref{sec:results-VLSM} showcase the true power of the framework developed in this paper: the possibility of rationally designing compliant walls, with properties optimized to suppress energetically important features of wall-bounded turbulent flows.  This optimization indicates that walls with negative damping, which lead to the wall-normal velocity and pressure being in phase at the wall, are required to suppress the NW cycle.  In addition to being consistent with heuristic arguments made in previous studies \citep{Xu2003}, perhaps this explains why DNS studies, carried out at low Reynolds numbers where the NW cycle is dominant, have met with limited success.  The optimization shows that walls with positive damping can effectively suppress modes resembling the VLSMs, suggesting that compliant walls may be better suited for high-Reynolds number applications.

Unfortunately, walls optimized to suppress specific types of resolvent modes or flow structures can have unanticipated detrimental effects elsewhere in spectral space, which could lead to a net deterioration in performance.  Hence, optimization for the NW cycle alone, which is often thought to be the determining factor in flow control, may not be sufficient to yield a global improvement in performance.  In particular, \S\ref{sec:results-Previous} and \S\ref{sec:results-Walls} indicate that there are sharp transitions in performance around the resonant frequency of the wall and that large slow-moving, spanwise-constant modes are particularly susceptible to being further amplified over compliant walls.  Such issues may be mitigated by designing walls that are slightly detuned and resonant away from the region of interest or have a specific wavenumber-frequency bandwidth.  For example, introducing the restoring effects of tension or stiffness would impose additional low-pass filters on wavenumber.  Of course, such low-pass filters would not eliminate the aforementioned sensitively to long, spanwise constant flow structures.  However, there may be other possibilities, including the use of finite-length panels.  The emergence of tunable \textit{metamaterials} may also offer significant scope for design.


\vspace{10pt}

This material is based on work supported by the Air Force Office of Scientific Research under awards FA9550-12-1-0469 (program manager Douglas Smith) and FA9550-14-1-0042 (program manager Gregg Abate).

\bibliographystyle{jfm}
\bibliography{2014-Luhar-CompliantSurface}

\begin{thebibliography}{47}
\expandafter\ifx\csname natexlab\endcsname\relax\def\natexlab#1{#1}\fi

\bibitem[Benjamin(1960)]{Benjamin1960}
{\sc Benjamin, T.~B.} 1960 Effects of a flexible boundary on hydrodynamic
  stability. {\em Journal of Fluid Mechanics\/} {\bf 9}, 513--532.

\bibitem[Benjamin(1963)]{Benjamin1963}
{\sc Benjamin, T.~B.} 1963 The threefold classification of unstable
  disturbances in flexible surfaces bounding inviscid flows. {\em Journal of
  Fluid Mechanics\/} {\bf 16}, 436--450.

\bibitem[Bourguignon {\em et~al.\/}(2014)Bourguignon, Tropp, Sharma \&
  McKeon]{Bourguignon2014}
{\sc Bourguignon, J.-L., Tropp, J.~A., Sharma, A.~S. \& McKeon, B.~J.} 2014
  Compact representation of wall-bounded turbulence using compressive sampling.
  {\em Physics of Fluids\/} {\bf 26}, 015109.

\bibitem[Bushnell {\em et~al.\/}(1977)Bushnell, Hefner \& Ash]{Bushnell1977}
{\sc Bushnell, D.~M., Hefner, J.~N. \& Ash, R.~L.} 1977 Effect of compliant
  wall motion on turbulent boundary layers. {\em Physics of Fluids\/} {\bf 20},
  S31--S48.

\bibitem[Carpenter(1990)]{Carpenter1990}
{\sc Carpenter, P.~W.} 1990 Status of transition delay using compliant walls.
  In {\em Viscous Drag Reduction in Boundary Layers\/} (ed. D.~M. Bushnell \&
  J.~N. Hefner), , vol. 123, pp. 79--113. Washington, {DC}: Progress in
  Aeronautics and Astronautics, AIAA.

\bibitem[Carpenter {\em et~al.\/}(2000)Carpenter, Davies \&
  Lucey]{Carpenter2000}
{\sc Carpenter, P.~W., Davies, C. \& Lucey, A.~D.} 2000 Hydrodynamics and
  compliant walls: does the dolphin have a secret? {\em Current Science\/} {\bf
  79}, 758--765.

\bibitem[Carpenter \& Garrad(1985)]{Carpenter1985}
{\sc Carpenter, P.~W. \& Garrad, A.~D.} 1985 The hydrodynamic stability of flow
  over {K}ramer-type compliant surfaces. {P}art 1. {T}ollmien-{S}chlichting
  instabilities. {\em Journal of Fluid Mechanics\/} {\bf 155}, 465--510.

\bibitem[Carpenter \& Garrad(1986)]{Carpenter1986}
{\sc Carpenter, P.~W. \& Garrad, A.~D.} 1986 The hydrodynamic stability of flow
  over {K}ramer-type compliant surfaces. {P}art 2. {F}low-induced surface
  instabilities. {\em Journal of Fluid Mechanics\/} {\bf 170}, 199--232.

\bibitem[Choi {\em et~al.\/}(1994)Choi, Moin \& Kim]{Choi1994}
{\sc Choi, H., Moin, P. \& Kim, J.} 1994 Active turbulence control for drag
  reduction in wall-bounded flows. {\em Journal of Fluid Mechanics\/} {\bf
  262}, 75--110.

\bibitem[Choi {\em et~al.\/}(1997)Choi, Yang, Clayton, Glover, Atlar, Semenov
  \& Kulik]{Choi1997}
{\sc Choi, K.~S., Yang, X., Clayton, B.~R., Glover, E.~J., Atlar, M., Semenov,
  B.~N. \& Kulik, V.~M.} 1997 Turbulent drag reduction using compliant
  surfaces. {\em Proceedings of the Royal Society A-Mathematical Physical and
  Engineering Sciences\/} {\bf 453}~(1965), 2229--2240.

\bibitem[Davies \& Carpenter(1997)]{Davies1997}
{\sc Davies, C. \& Carpenter, P.~W.} 1997 Instabilities in a plane channel flow
  between compliant walls. {\em Journal of Fluid Mechanics\/} {\bf 352},
  205--243.

\bibitem[Dixon {\em et~al.\/}(1994)Dixon, Lucey \& Carpenter]{Dixon1994}
{\sc Dixon, A.~E., Lucey, A.~D. \& Carpenter, P.~W.} 1994 Optimization of
  viscoelastic compliant walls for transition delay. {\em AIAA journal\/} {\bf
  32}~(2), 256--267.

\bibitem[Duncan(1986)]{Duncan1986}
{\sc Duncan, J.~H.} 1986 The response of an imcompressible viscoelastic coating
  to pressure fluctuations in a turbulent boundary layer. {\em Journal of Fluid
  Mechanics\/} {\bf 171}, 339--363.

\bibitem[Endo \& Himeno(2002)]{Endo2002}
{\sc Endo, T. \& Himeno, R.} 2002 Direct numerical simulation of turbulent flow
  over a compliant surface. {\em Journal of Turbulence\/} {\bf 3}, 1--10.

\bibitem[Fukagata {\em et~al.\/}(2002)Fukagata, Iwamoto \&
  Kasagi]{Fukagata2002}
{\sc Fukagata, K., Iwamoto, K. \& Kasagi, N.} 2002 Contribution of {R}eynolds
  stress distribution to the skin friction in wall-bounded flows. {\em Physics
  of Fluids\/} {\bf 14}~(11), 73--76.

\bibitem[Fukagata {\em et~al.\/}(2008)Fukagata, Kern, Chatelain, Koumoutsakos
  \& Kasagi]{Fukagata2008}
{\sc Fukagata, K., Kern, S., Chatelain, P., Koumoutsakos, P. \& Kasagi, N.}
  2008 Evolutionary optimization of an anisotropic compliant surface for
  turbulent friction drag reduction. {\em Journal of Turbulence\/} {\bf
  9}~(35), 1--17.

\bibitem[{Gad-el-Hak}(1986)]{GadelHak1986}
{\sc {Gad-el-Hak}, M.} 1986 The response of elastic and viscoelastic surfaces
  to a turbulent boundary layer. {\em Journal of Applied Mechanics\/} {\bf
  53}~(1), 206--212.

\bibitem[{Gad-el-Hak}(2000)]{GadelHak2000}
{\sc {Gad-el-Hak}, M.} 2000 {\em Flow control: passive, active, and reactive
  flow management\/}. Cambridge University Press, London, UK.

\bibitem[{Gad-el-Hak} {\em et~al.\/}(1984){Gad-el-Hak}, Blackwelder \&
  Riley]{GadelHak1984}
{\sc {Gad-el-Hak}, M., Blackwelder, R.~F. \& Riley, J.~J.} 1984 On the
  interaction of compliant coatings with boundary-layer flows. {\em Journal of
  Fluid Mechanics\/} {\bf 140}, 257--280.

\bibitem[Hooke \& Jeeves(1961)]{Hooke1961}
{\sc Hooke, R. \& Jeeves, T.~A.} 1961 Direct search solution of numerical and
  statistical problems. {\em Journal of the ACM\/} {\bf 8}~(2), 212--229.

\bibitem[Hoyas \& Jimenez(2006)]{Hoyas2006}
{\sc Hoyas, S. \& Jimenez, J.} 2006 Scaling of the velocity fluctuations in
  turbulent channels up to {R}e$_\tau$ = 2003. {\em Physics of Fluids\/} {\bf
  18}~(011702).

\bibitem[Kim \& Choi(2014)]{Kim2014}
{\sc Kim, E. \& Choi, H.} 2014 Space-time characteristics of a compliant wall
  in a turbulent channel flow. {\em Journal of Fluid Mechanics\/} {\bf 756},
  30--53.

\bibitem[Kireiko(1990)]{Kireiko1990}
{\sc Kireiko, G.~V.} 1990 Interaction of wall turbulence with a compliant
  surface. {\em Fluid Dynamics\/} {\bf 25}~(4), 550--554.

\bibitem[Koumoutsakos(1999)]{Koumoutsakos1999}
{\sc Koumoutsakos, P.} 1999 Vorticity flux control for a turbulent channel
  flow. {\em Physics of Fluids\/} {\bf 11}, 248.

\bibitem[Kramer(1961)]{Kramer1961}
{\sc Kramer, M.~O.} 1961 The dolphin's secret. {\em Journal of the American
  Society for Naval Engineers\/} {\bf 73}, 103--108.

\bibitem[Landahl(1962)]{Landahl1962}
{\sc Landahl, M.~T.} 1962 On the stability of laminar incompressible boundary
  layer flow over a flexible surface. {\em Journal of Fluid Mechanics\/} {\bf
  13}, 609--632.

\bibitem[Lee {\em et~al.\/}(1993)Lee, Fisher \& Schwarz]{Lee1993}
{\sc Lee, T., Fisher, M. \& Schwarz, W.~H.} 1993 Investigation of the stable
  interaction of a passive compliant surface with a turbulent boundary layer.
  {\em Journal of Fluid Mechanics\/} {\bf 257}, 373--401.

\bibitem[Lucey \& Carpenter(1995)]{Lucey1995}
{\sc Lucey, A.~D. \& Carpenter, P.~W.} 1995 Boundary layer instability over
  compliant walls: comparison between theory and experiment. {\em Physics of
  Fluids\/} {\bf 7}, 2355--2363.

\bibitem[Luhar {\em et~al.\/}(2014{\natexlab{{\em a\/}}})Luhar, Sharma \&
  McKeon]{Luhar2014b}
{\sc Luhar, M., Sharma, A.~S. \& McKeon, B.~J.} 2014{\natexlab{{\em a\/}}} On
  the structure and origin of pressure fluctuations in wall turbulence:
  predictions based on the resolvent analysis. {\em Journal of Fluid
  Mechanics\/} {\bf 751}, 38--70.

\bibitem[Luhar {\em et~al.\/}(2014{\natexlab{{\em b\/}}})Luhar, Sharma \&
  McKeon]{Luhar2014a}
{\sc Luhar, M., Sharma, A.~S. \& McKeon, B.~J.} 2014{\natexlab{{\em b\/}}}
  Opposition control within the resolvent analysis framework. {\em Journal of
  Fluid Mechanics\/} {\bf 749}, 597--626.

\bibitem[Marusic {\em et~al.\/}(2010)Marusic, Mathis \& Hutchins]{Marusic2010}
{\sc Marusic, I., Mathis, R. \& Hutchins, N.} 2010 Predictive model for
  wall-bounded turbulent flow. {\em Science\/} {\bf 329}~(5988), 193--196.

\bibitem[McKeon {\em et~al.\/}(2013)McKeon, Jacobi \& Sharma]{McKeon2013}
{\sc McKeon, B.~J., Jacobi, I. \& Sharma, A.~S.} 2013 Experimental manipulation
  of wall turbulence: a systems approach. {\em Physics of Fluids\/} {\bf 25},
  031301.

\bibitem[McKeon \& Sharma(2010)]{McKeon2010}
{\sc McKeon, B.~J. \& Sharma, A.~S.} 2010 A critical-layer framework for
  turbulent pipe flow. {\em Journal of Fluid Mechanics\/} {\bf 658}, 336--382.

\bibitem[Moarref \& Jovanovic(2012)]{Moarref2012}
{\sc Moarref, R. \& Jovanovic, M.} 2012 Model-based design of transverse wall
  oscillations for turbulent drag reduction. {\em Journal of Fluid Mechanics\/}
  {\bf 707}, 205--240.

\bibitem[Moarref {\em et~al.\/}(2013)Moarref, Sharma, Tropp \&
  McKeon]{Moarref2013}
{\sc Moarref, R., Sharma, A.~S., Tropp, J.~A. \& McKeon, B.~J.} 2013
  Model-based scaling and prediction of the streamwise energy intensity in
  high-{R}eynolds number turbulent channels. {\em Journal of Fluid Mechanics,
  accepted\/} ArXiv:1302.1594 [physics.flu-dyn].

\bibitem[Monty \& Chong(2009)]{Monty2009a}
{\sc Monty, J.~P. \& Chong, M.~S.} 2009 Turbulent channel flow: comparison of
  streamwise velocity data from experiments and direct numerical simulation.
  {\em Journal of Fluid Mechanics\/} {\bf 633}, 461--474.

\bibitem[Nakanishi {\em et~al.\/}(2012)Nakanishi, Mamori \&
  Fukagata]{Nakanishi2012}
{\sc Nakanishi, R., Mamori, H. \& Fukagata, K.} 2012 Relaminarization of
  turbulent channel flow using traveling wave-like wall deformation. {\em
  International Journal of Heat and Fluid Flow\/} {\bf 35}, 152--159.

\bibitem[Rempfer {\em et~al.\/}(2001)Rempfer, Blossey, Parsons \&
  Lumley]{Rempfer2001}
{\sc Rempfer, D., Blossey, P., Parsons, L. \& Lumley, J.} 2001 Low-dimensional
  dynamical model of a turbulent boundary layer over a compliant surface:
  preliminary results. In {\em Fluid Mechanics and the Environment: Dynamical
  Approaches\/}, pp. 267--283.

\bibitem[Reynolds \& Tiederman(1967)]{Reynolds1967}
{\sc Reynolds, W.~C. \& Tiederman, W.~G.} 1967 Stability of turbulent channel
  flow with application to {M}alkus's theory. {\em Journal of Fluid
  Mechanics\/} {\bf 27}, 253--272.

\bibitem[Riley {\em et~al.\/}(1988)Riley, {Gad-el-Hak} \& Metcalfe]{Riley1988}
{\sc Riley, J.~J., {Gad-el-Hak}, M. \& Metcalfe, R.~W.} 1988 Compliant
  coatings. {\em Annual Review of Fluid Mechanics\/} {\bf 20}, 393--420.

\bibitem[Robinson(1991)]{Robinson1991}
{\sc Robinson, S.~K.} 1991 Coherent motions in the turbulent boundary-layer.
  {\em Annual Review of Fluid Mechanics\/} {\bf 23}, 601--639.

\bibitem[Sen \& Arora(1988)]{Sen1988}
{\sc Sen, P.~K. \& Arora, D.~S.} 1988 On the stability of laminar
  boundary-layer flow over a flat plate with a compliant surface. {\em Journal
  of Fluid Mechanics\/} {\bf 197}, 201--240.

\bibitem[Sharma \& McKeon(2013)]{Sharma2013}
{\sc Sharma, A.~S. \& McKeon, B.~J.} 2013 On coherent structure in wall
  turbulence. {\em Journal of Fluid Mechanics\/} {\bf 728}, 196--238.

\bibitem[Smits {\em et~al.\/}(2011)Smits, Monty, Hultmark, Bailey, Hutchins \&
  Marusic]{Smits2011}
{\sc Smits, A.~J., Monty, J., Hultmark, M., Bailey, S. C.~C., Hutchins, N. \&
  Marusic, I.} 2011 Spatial resolution correction for wall-bounded turbulence
  measurements. {\em Journal of Fluid Mechanics\/} {\bf 676}, 41--53.

\bibitem[Trefethen(2000)]{Trefethen2000}
{\sc Trefethen, L.~N.} 2000 {\em Spectral methods in {MATLAB}\/}. Society for
  Industrial and Applied Mathematics, USA.

\bibitem[Weideman \& Reddy(2000)]{Weideman2000}
{\sc Weideman, J. A.~C. \& Reddy, S.~C.} 2000 A {MATLAB} differentiation matrix
  suite. {\em ACM Transactions on Mathematical Software\/} {\bf 26}, 465--519.

\bibitem[Xu {\em et~al.\/}(2003)Xu, Rempfer \& Lumley]{Xu2003}
{\sc Xu, S., Rempfer, D. \& Lumley, J.} 2003 Turbulence over a compliant
  surface: numerical simulation and analysis. {\em Journal of Fluid
  Mechanics\/} {\bf 478}, 11--34.

\end{thebibliography}
\end{document}